\documentclass[%
 amsmath,amssymb,
 aps,
 prd,
twocolumn, superscriptaddress
]{revtex4-1}

\usepackage{float}
\usepackage[normalem]{ulem}

\usepackage{graphicx}
\usepackage{dcolumn}
\usepackage{bm}
\usepackage{hyperref}

\usepackage{xcolor}

\usepackage{color}

\begin{document}

\preprint{APS/123-QED} 

\title{Charged quark stars in $f(R,T)$ gravity }

\author{Juan M. Z. Pretel}
 \email{juanzarate@cbpf.br}
 \affiliation{Centro Brasileiro de Pesquisas F{\'i}sicas, Rua Dr. Xavier Sigaud, 150 URCA, Rio de Janeiro CEP 22290-180, RJ, Brazil
}

\author{Takol Tangphati} 
 \email{takoltang@gmail.com}
 \affiliation{Theoretical and Computational Physics Group, \\
 Theoretical and Computational Science Center (TaCS), Faculty of Science, \\
 King Mongkut's University of Technology Thonburi, 126 Prachauthid Rd., Bangkok 10140, Thailand
}

\author{Ayan Banerjee} 
 \email{ayanbanerjeemath@gmail.com}
 \affiliation{Astrophysics and Cosmology Research Unit, School of Mathematics, Statistics and Computer Science, University of KwaZulu--Natal, Private Bag X54001, Durban 4000, South Africa
}

\author{Anirudh Pradhan}
 \email{pradhan.anirudh@gmail.com}
 \affiliation{Centre for Cosmology, Astrophysics and Space Science, GLA University, Mathura-281 406, Uttar Pradesh, India
}

\date{\today}

\begin{abstract}
Recent advances in nuclear theory combined with new astrophysical observations have led to the need for specific theoretical models that actually apply to phenomena on dense-matter physics. At the same time, quantum chromodynamics (QCD) predicts the existence of non-nucleonic degrees of freedom at high densities in neutron-star matter, such as quark matter. Within a confining quark matter model, which consists of homogeneous, neutral 3-flavor interacting quark matter with $\mathcal{O}(m_s^4)$ corrections, we study the structure of compact stars  made of a charged perfect fluid in the context of $f(R,T)$ gravity. The system of differential equations that describe the structure of charged compact stars have been derived and solved numerically for a gravity model with $f(R,T)= R+ 2\beta T$. For simplicity, we assume that the charge density is proportional to the energy density, namely, $\rho_{\rm ch} = \alpha \rho$. It is demonstrated that matter-geometry coupling constant $\beta$ and the charge parameter $\alpha$ affect the total gravitational mass and the radius of the star.

\end{abstract}

\maketitle


\section{Introduction}

Despite being over 100 years old, Einstein's General Relativity (GR) is the most successful theory of modern physics to explain gravitational phenomena. In fact, the predictions of GR have passed all observational/experimental tests both in the weak field limit and more recently for strongly self-gravitating bodies in pulsar binary systems. Nonetheless, there are many issues that cannot be explained through GR both from a theoretical and observational point of view \cite{Coley:2018mzr} and thereby a number of alternative models have been proposed. In particular, instead of introducing unknown fluids like dark energy, one can propose alternative theories of gravity extending the Einstein theory. This has been an active research area in the last few years, and in the present work we adopt a similar path.

In this direction, the simplest model of modified gravity is $f(R)$ gravity \cite{Sotiriou:2008rp, DeFelice:2010aj}, where $f(R)$ is a generic function of the Ricci scalar $R$. This modification comes into the game by replacing the Einstein-Hilbert Lagrangian by an arbitrary function $f(R)$. Such gravity theory can explain the late-time acceleration of the cosmic expansion without the need of introducing a dark energy field (see e.g. Refs.~\cite{Nojiri:2010wj, Nojiri:2017ncd}). A generalization of $f(R)$ gravity was proposed in \cite{Capozziello:2014bqa},  where additional degrees of freedom related to curvature invariants and scalar fields have been considered in Extended Theories of Gravity (ETG). Interestingly, these new degrees of freedom can be recast as effective fluids that differ in meaning with respect to the standard matter fluids generally adopted as sources of the  field equations. Furthermore, some authors have shown that ETG may be used to account for the accelerated expansion of the Universe without any explicit additional matter component \cite{Bertolami:2010cw}.

Motivated by these results, the inclusion of matter terms in the gravitational action was further explored in $f(R, T)$ gravity \cite{Harko2011}, where the modified Lagrangian is an arbitrary function of Ricci scalar and the trace of the energy-momentum tensor $T$. Within this context, many authors have explored cosmological solutions by choosing several functional forms of $f$, such as cosmological solutions based on a homogeneous and isotropic space-time through a phase-space analysis \cite{Shabani:2013djy}, stability analysis using energy conditions \cite{Alvarenga:2012bt, Sharif:2012ce} and thermodynamics aspects and its larger consequences have been discussed in \cite{Sharif:2012zzd, Jamil:2012pf, Sharif:2013ffa}. Meanwhile, cosmological and solar system consequences have been discussed in \cite{Shabani:2014xvi}. The authors of Ref.~\cite{Xu:2016rdf} studied the quantum cosmology of $f(R, T)$ gravity.

Although the viability of some $f(R,T)$ models in the cosmological scenario has already been discussed \cite{Velten2017, Godani2019, Gamonal2020}, these theories can also be used as a test-bed for dealing with the strong-gravity regime such as compact stars. The matter inside them is compressed at sufficiently high densities and the ``escape velocity'' near the surface is close to the speed of light. Therefore, it is not easy to produce similar environments in a terrestrial laboratory. Theoretical models can be uniquely studied considering an equation of state (EoS) where different kinds of EoS can be taken into account. Besides that, the most accurate mass-radius relations from radio pulsars and some other accretion-powered X-ray sources have set rigid constraints on various EoS at strong-density regimes.

In $f(R, T)$ gravity, the stellar equilibrium configurations of compact stars have been widely investigated (see for instance Refs.~\cite{Moraes2016, Das:2016mxq, Deb:2017rhd, Biswas:2018inc, Lobato2020, Pretel2021} and references therein). Within the same context, Pretel and collaborators have studied the radial stability of compact stars and showed that the traditional stellar stability criteria still hold \cite{Pretel:2020oae}. In considering these theories, several stellar models with isotropic pressure have been proposed in \cite{Hansraj:2018cmj, Maurya:2019msr}. Some works analyzed the stability criteria by imposing some constraints on the EoS \cite{Deb:2017rhc, Deb:2018sgt, Maurya:2019sfm, Yousaf:2018jkb}. In Ref.~\cite{Hansraj:2018jzb}, the dynamical behavior of the Tolman metrics has been studied. Moreover, there have been constructed models of Mazur-Mottola gravastar \cite{Das:2017rhi}, wormholes \cite{Moraes:2017mir, Banerjee:2019wjj, Elizalde:2018arz}, and quark stars \cite{Deb:2018gzt, Sharif:2018khl, Tangphati:2022mur}. Indeed, a maximum mass of $M_{\text{TOV}}$ $> 2.5 M_{\odot}$ \cite{Tangphati:2022arm} for static QS has been obtained based on a careful investigation by applying a quark EoS to simultaneously analyze the GW190814 data.

It is a general consensus that the physics governing the structure of stars does not process a large amount of charge in nature \cite{Eddington}. However, several researchers disagree with this argument (see e.g.~Refs.~\cite{Olson, Olson1} for review). According to them, the matter acquires a large amount of electric charge during the gravitational collapse or during an accretion process onto NSs if accretion produces luminosity close to the Eddington limit \cite{Shvartsman}. In order to establish this gravitational phenomenon, it was shown that electrically charged quark matter may lead to huge electric fields on the order of $10^{18}$ V/cm \cite{alcock86:a, alcock88:a}. In addition, Ray \textit{et al.} \cite{Ray:2003gt} examined the possibility of any significant changes on the mass-radius relation in presence of electric fields carried by compact stars. Very recently, this finding is verified in a few other models of charged stars \cite{Panotopoulos:2021cxu, Lemos:2014lza, Arbanil:2017huq}. From the above reasons, our goal is to explore charged compact stars in hydrostatic equilibrium as astrophysical tests of  $f(R,T)$ gravity in a strong-gravity regime.

This paper is organized as follows: In Section \ref{sec2}, we give a brief review of the $f(R,T)$ gravity and derive the field equations in the presence of electromagnetic field. In the same section, we consider a spherically symmetric metric and derive the modified Tolman-Oppenheimer-Volkoff (TOV) equations. In Section \ref{sec3} we present an overview of a QCD motivated EoS and the charge distribution profile. In Section \ref{sec4}, we briefly describe the numerical method employed to determine the mass-radius relations of QSs for two different sets of parameters. This section is also devoted to study a relative difference between the maximum-mass values in GR and $f(R,T)$ gravity. Finally, in Section \ref{sec5}, we draw conclusions from our results.


\section{Stellar structure equations}\label{sec2}

\subsection{Field equations in $f(R,T)$ gravity}

Harko and collaborators \cite{Harko2011} constructed a more general formulation of $f(R)$ theories of gravity by means of an arbitrary function $f(R,T)$, where $R$ and $T$ are the scalar curvature and the trace of the energy-momentum tensor $T_{\mu\nu}$, respectively. To describe electrically charged compact stars, the gravitational action in $f(R,T)$ is supplemented by the standard matter and electromagnetic field. Thus, in our study, the total action is given by
\begin{align}\label{ActionEq}
    S &= S_g + S_m + S_e  \nonumber  \\
    &= \int d^{4}x  \sqrt{-g} \left[\frac{1}{16\pi} f(R,T) + \mathcal{L}_m + \mathcal{L}_e \right] ,
\end{align}
where $g$ is the determinant of the metric tensor $g_{\mu\nu}$ and $\mathcal{L}_m$ denotes the matter Lagrangian density. Besides, the Lagrangian density of electromagnetic field in curved spacetime takes the form
\begin{equation}
    \mathcal{L}_e = j^\mu A_\mu - \frac{1}{16\pi}F_{\mu\nu}F_{\lambda\sigma}g^{\mu\lambda}g^{\nu\sigma} ,
\end{equation}
where $j^{\mu} = \rho_{\rm ch}u^\mu$ is the four-current density with $\rho_{\rm ch}$ being the electric charge density and $u^\mu$ the four-velocity of the fluid, $F_{\mu\nu}= \nabla_\mu A_\nu - \nabla_\nu A_\mu$ is the electromagnetic field strength tensor, $\nabla_\mu$ the covariant derivative and $A_\mu$ is the electromagnetic four-potential. 

As in the pure GR case \cite{Negreiros2009, Arbanil2013}, we consider that the total energy-momentum tensor for the charged fluid is the sum of two contributions, namely $T_{\mu\nu}= \mathcal{M}_{\mu\nu} + \mathcal{E}_{\mu\nu}$, where $\mathcal{M}_{\mu\nu}$ and $\mathcal{E}_{\mu\nu}$ are the matter and electromagnetic energy-momentum tensors, respectively. The latter is given by
\begin{equation}\label{EMTensor2}
    \mathcal{E}_{\mu\nu} = \frac{-2}{\sqrt{-g}}\frac{\delta S_e}{\delta g^{\mu\nu}} = \frac{1}{4\pi} \left[ F_{\mu\lambda}F_\nu^{\ \lambda} - \frac{1}{4}g_{\mu\nu}F_{\lambda\sigma}F^{\lambda\sigma} \right] ,
\end{equation}
which is traceless. This implies that $T \equiv g^{\mu\nu}T_{\mu\nu}= \mathcal{M}$ and hence $\delta T = \mathcal{M}_{\mu\nu}\delta g^{\mu\nu} + g^{\mu\nu}\delta\mathcal{M}_{\mu\nu}$. 

Consequently, the variation of the gravitational action with respect to the inverse metric leads to 
\begin{align}\label{VariaSm}
    \delta S_g =& \int \frac{\sqrt{-g}}{16\pi} \left[ f_RR_{\mu\nu} - \frac{1}{2}g_{\mu\nu}f + \left( g_{\mu\nu}\square - \nabla_\mu\nabla_\nu \right)f_R  \right.  \nonumber  \\ 
    &\hspace{1.5cm} + f_T\left( \mathcal{M}_{\mu\nu} + \Theta_{\mu\nu} \right) \bigg]\delta g^{\mu\nu}d^4x ,
\end{align}
where $f_R \equiv \partial f(R, T)/\partial R$, $f_T \equiv \partial f(R, T)/\partial T$, $\square \equiv \nabla_\mu\nabla^\mu$ is the d'Alembert operator, $\mathcal{M}_{\mu\nu}$ is the ordinary matter energy-momentum tensor written as 
\begin{equation}\label{EMTensor1}
    \mathcal{M}_{\mu\nu} = \frac{-2}{\sqrt{-g}}\frac{\delta S_m}{\delta g^{\mu\nu}} = g_{\mu\nu}\mathcal{L}_m - 2\frac{\partial\mathcal{L}_m}{\partial g^{\mu\nu}} ,
\end{equation}
and the tensor $\Theta_{\mu\nu}$ is defined as 
\begin{align}
    \Theta_{\mu\nu} &= g^{\alpha\beta}\frac{\delta\mathcal{M}_{\alpha\beta}}{\delta g^{\mu\nu}}  \nonumber  \\
    &= -2\mathcal{M}_{\mu\nu} + g_{\mu\nu}\mathcal{L}_m - 2g^{\alpha\beta}\frac{\partial^2\mathcal{L}_m}{\partial g^{\mu\nu} \partial g^{\alpha\beta}} .  \label{ThetaEq}
\end{align}

From the variational principle $\delta S= 0$ and in view of Eqs.~(\ref{EMTensor2})-(\ref{EMTensor1}), we obtain the following field equations 
\begin{align}\label{FieldEq}
   &f_R R_{\mu\nu} - \dfrac{1}{2}g_{\mu\nu}f + g_{\mu\nu}\square f_R - \nabla_\mu\nabla_\nu f_R  \nonumber  \\
   &\hspace{1cm} = 8\pi(\mathcal{M}_{\mu\nu} + \mathcal{E}_{\mu\nu}) - f_T(\mathcal{M}_{\mu\nu} + \Theta_{\mu\nu}) ,
\end{align}
which are fourth-order differential equations in the metric functions, and the standard Einstein-Maxwell equation is retrieved when $f(R,T)= R$. Similar to $f(R)$ gravity, the Ricci scalar is a dynamical quantity in $f(R,T)$ theories. In other words, the scalar curvature is governed by the trace of the field equations (\ref{FieldEq}), that is, 
\begin{equation}\label{TraceEq}
    3\square f_R + Rf_R - 2f = 8\pi \mathcal{M} - f_T(\mathcal{M} + \Theta) ,
\end{equation}
and the covariant divergence yields
\begin{align}\label{CovaDiverEq}
    \nabla^\mu\mathcal{M}_{\mu\nu} =&\ \frac{f_T}{8\pi - f_T}\left[ \left( \mathcal{M}_{\mu\nu} + \Theta_{\mu\nu} \right)\nabla^\mu\ln f_T + \nabla^\mu\Theta_{\mu\nu} \right. \nonumber  \\
    &\left. -\frac{1}{2}g_{\mu\nu}\nabla^\mu T - \frac{8\pi}{f_T}\nabla^\mu\mathcal{E}_{\mu\nu} \right] .
\end{align}

In order to continue our work, here it is necessary to specify the particular $f(R,T)$ gravity model. Following Ref.~\cite{Harko2011}, the simplest model is given by $f(R,T) = R+ 2\beta T$, where $\beta$ is a coupling parameter. Furthermore, as in Refs.~\cite{Moraes2016, Lobato2020, Pretel:2020oae, Tangphati:2022arm, Tangphati:2022mur}, the matter Lagrangian density can be taken as $\mathcal{L}_m = p$. Therefore, we have $\Theta_{\mu\nu} = -2\mathcal{M}_{\mu\nu} + pg_{\mu\nu}$, $\Theta = -2\mathcal{M}+ 4p$, and Eqs.~(\ref{FieldEq})-(\ref{CovaDiverEq}) assume the following form 
\begin{align}
    &G_{\mu\nu} = 8\pi T_{\mu\nu} + \beta\mathcal{M}g_{\mu\nu} + 2\beta\left( \mathcal{M}_{\mu\nu} - pg_{\mu\nu} \right) ,  \label{FieldEqSimp}  \\  
    &R = -8\pi \mathcal{M} + 2\beta\left( -3\mathcal{M} + 4p \right) ,  \label{TraceEqSimp}  \\
    &\nabla^\mu\mathcal{M}_{\mu\nu} = \frac{2\beta}{8\pi + 2\beta}\left[ \nabla_\nu \left(p- \frac{\mathcal{M}}{2} \right) - \frac{4\pi}{\beta}\nabla^\mu\mathcal{E}_{\mu\nu} \right] ,  \label{CovaDiverEqSimp}
\end{align}
where $G_{\mu\nu}$ is the usual Einstein tensor. It is evident that the general relativistic equations are obtained when $\beta= 0$.

\subsection{Modified TOV equations}

To explore non-spinning isolated compact stars, we assume that the spacetime can be described by the usual spherically symmetric metric
\begin{equation}\label{MetricEq}
    ds^2 = -e^{2\psi}dt^2 + e^{2\lambda}dr^2 + r^2(d\theta^2 + \sin^2\theta d\phi^2) ,
\end{equation}
which gives a distance rule $ds^2 = g_{\mu\nu}dx^\mu dx^\nu$ with $x^\mu = (t,r,\theta,\phi)$ being the components of the four-position vector and the metric functions $\psi$ and $\lambda$ depend only on the radial coordinate $r$. Moreover, we model the dense matter by a isotropic perfect fluid whose energy-momentum tensor is given by $\mathcal{M}_\mu^{\ \nu} = {\rm diag}(-\rho, p, p, p)$, where $\rho$ is the energy density and $p$ is the pressure of the fluid.

The electromagnetic strength tensor must satisfy the Maxwell equations, namely
\begin{align}
    &\frac{1}{\sqrt{-g}} \partial_\mu\left( \sqrt{-g}F^{\mu\nu} \right) = -4\pi j^\nu ,  \label{MEq1}  \\
    &\nabla_\sigma F_{\mu\nu} + \nabla_\mu F_{\nu\sigma} + \nabla_\nu F_{\sigma\mu} = 0 ,  \label{MEq2}
\end{align}
which lead to the expression $\nabla_\mu\mathcal{E}^{\mu\nu} = j_\lambda F^{\lambda\nu}$. Since the only non-zero component of the strength tensor is $F^{01} = -F^{10}$, the Maxwell equation (\ref{MEq1}) becomes
\begin{equation}
    F^{01} = \frac{q}{r^2}e^{-\psi - \lambda} , 
\end{equation}
where the electric charge function $q(r)$ is given by 
\begin{equation}\label{ChargeEq}
    q(r) = 4\pi\int _0^r \bar{r}^2\rho_{\rm ch}(\bar{r})e^{\lambda(\bar{r})}d\bar{r} .
\end{equation}

By substituting these quantities into Eq.~(\ref{FieldEqSimp}), we obtain the non-vanishing components of the field equations, namely
\begin{equation}\label{FieldEq1}
    \frac{1}{r^2}\frac{d}{dr}\left( re^{-2\lambda} \right) - \frac{1}{r^2} = -8\pi\left[ \rho + \frac{q^2}{8\pi r^4} \right] + \beta(-3\rho + p) ,
\end{equation}
\begin{equation}\label{FieldEq2}
    \frac{1}{e^{2\lambda}}\left( \frac{2}{r}\psi'+ \frac{1}{r^2} \right) - \frac{1}{r^2} = 8\pi\left[ p - \frac{q^2}{8\pi r^4} \right] + \beta(-\rho+ 3p) ,
\end{equation}
\begin{align}\label{FielEq3}
    &\frac{1}{e^{2\lambda}}\left[ \psi''+ \psi'^2 - \psi'\lambda' + \frac{1}{r}(\psi' - \lambda') \right]  \nonumber  \\
    &\hspace{1.2cm}= 8\pi\left[ p + \frac{q^2}{8\pi r^4} \right] + \beta(-\rho+ 3p) ,
\end{align}
where the prime indicates differentiation with respect to $r$. The non-conservation of the energy-momentum tensor (\ref{CovaDiverEqSimp}) reduces to
\begin{equation}\label{NonConservEq}
    p' = -\left( \frac{\rho + p}{1+a} \right)\psi' + \left( \frac{1-2a}{1+a} \right)\frac{\rho_{\rm ch}e^\lambda q}{r^2} + \frac{a\rho'}{1+a} ,
\end{equation}
where we have defined $a \equiv \beta/(8\pi + 2\beta)$. 

Here it is pertinent to introduce a mass function that allows us to determine the gravitational mass of a charged star within the context of $f(R,T)= R+ 2\beta T$ gravity. To do so, we rewrite Eq.~(\ref{FieldEq1}) as 
\begin{align}
    \frac{d}{dr}\left( re^{-2\lambda} \right) = 1 &- 2\left[ 4\pi r^2\rho + \frac{\beta r^2}{2}(3\rho - p) + \frac{q}{r}\frac{dq}{dr} \right]  \nonumber  \\
    &+ \frac{d}{dr}\left( \frac{q^2}{r} \right) ,
\end{align}
and after integrating the last expression, one obtains
\begin{equation}\label{ExpLambda}
    e^{-2\lambda} = 1 - \frac{2m}{r} + \frac{q^2}{r^2} ,
\end{equation}
where the mass function $m(r)$ can be interpreted as the total gravitational mass contained in a charged sphere of radius $r$, 
\begin{equation}\label{MassFunc}
    m = 4\pi\int r^2\rho dr + \frac{\beta}{2}\int r^2(3\rho- p)dr + \int\frac{qq'}{r}dr .
\end{equation}

This means that the total mass of the star is the contribution of three terms: energy density of standard matter in the first integral, gravitational modification via $2\beta T$, and electric charge in the third integral. For $\beta= 0$, we recover the conventional mass function of a charged perfect fluid in Einstein gravity \cite{Negreiros2009, Arbanil2013}. Furthermore, in the uncharged case, the above equation reduces to the expression given in Ref.~\cite{Pretel:2020oae}. 

In view of Eq.~(\ref{ExpLambda}), the 11-component of the field equations (\ref{FieldEq2}) becomes 
\begin{equation}\label{PsiPrimeEq}
    \psi' = \left[ 4\pi rp + \frac{m}{r^2} - \frac{\beta r}{2}(\rho - 3p) - \frac{q^2}{r^3} \right]e^{2\lambda} . 
\end{equation}
Consequently, from Eqs.~(\ref{ChargeEq}), (\ref{NonConservEq}), (\ref{MassFunc}) and (\ref{PsiPrimeEq}), the hydrostatic equilibrium of a charged compact star within the framework of $f(R,T) = R+ 2\beta T$ gravity model is governed by the modified TOV equations 
\begin{align}
    \frac{dq}{dr} &= 4\pi r^2\rho_{\rm ch}e^\lambda ,  \label{TOV1}  \\  
    \frac{dm}{dr} &= 4\pi r^2\rho + \frac{\beta r^2}{2}(3\rho - p) + \frac{q}{r}\frac{dq}{dr},  \label{TOV2}  \\
    \frac{dp}{dr} &= -\frac{\rho+ p}{1+ a}\left[ 4\pi rp + \frac{m}{r^2} - \frac{\beta r}{2}(\rho - 3p) - \frac{q^2}{r^3} \right]e^{2\lambda}  \nonumber  \\
    &\hspace{0.5cm} + \left( \frac{1-2a}{1+a} \right)\frac{q}{4\pi r^4}\frac{dq}{dr} + \frac{a}{1+ a} \frac{d\rho}{dr} ,  \label{TOV3}  \\
    \frac{d\psi}{dr} &= -\frac{1+a}{\rho +p}\frac{dp}{dr} + \left( \frac{1-2a}{\rho +p} \right)\frac{q}{4\pi r^4}\frac{dq}{dr} + \frac{a}{\rho +p}\frac{d\rho}{dr} .  \label{TOV4}
\end{align}

This is a set of four first-order coupled ordinary differential equations for the six unknown variables $q$, $\rho_{\rm ch}$, $m$, $\rho$, $p$ and $\psi$. The metric function $\lambda$ can be calculated from Eq.~(\ref{ExpLambda}). As expected, when the electric charge vanishes, the system of equations (\ref{TOV1})-(\ref{TOV4}) reduces to the uncharged modified TOV equations obtained by Pretel \textit{et al.} \cite{Pretel:2020oae}.

Since we have more variables than differential equations, it is convenient to adopt an equation of state $p= p(\rho)$ and a charge distribution in the form $\rho_{\rm ch} = \rho_{\rm ch}(\rho)$. Thus, the number of unknown functions is reduced to four, and we can close the system of equations. To ensure regularity of spacetime geometry, we have to specify the following initial conditions at the center of the star
\begin{align}\label{BC1}
    q(0) &= 0,   &   m(0) &= 0 ,   &   \rho(0) &= \rho_c ,
\end{align}
where $\rho_c$ is the central energy density, and we integrate outwards up to the pressure vanishes. Namely, the stellar surface is determined when $p(r_{\rm sur})= 0$, where $r_{\rm sur}$ is the radius of the star. 

In addition, outside the compact star $\rho= p =0$ and in view of Eq.~(\ref{TraceEqSimp}), we obtain $R=0$. This means that exterior spacetime is still described by the standard Reissner-Nordstr{\"o}m exterior solution as in GR, and hence the continuity of the metric at the surface imposes a boundary condition for the differential equation (\ref{TOV4}), namely
\begin{equation}
    \psi(r_{\rm sur}) = \frac{1}{2}\ln\left[ 1 - \frac{2M}{r_{\rm sur}} + \frac{Q^2}{r_{\rm sur}^2} \right] ,
\end{equation}
where $m(r_{\rm sur}) \equiv M$ and $q(r_{\rm sur}) \equiv Q$ are the total mass and total charge determined at the surface, respectively. However, it is worth emphasizing that for non-linear functions in $R$ the Reissner-Nordstr{\"o}m metric is invalid because the scalar curvature does not vanish in the outer region of a compact star. As a matter of fact, see Ref.~\cite{Pretel2022} for a discussion about charged quark stars in quadratic $f(R)$ gravity.


\section{Equation of state and charge distribution} \label{sec3}

A core-collapse supernova explosion occurs in the last stage of giant stars at the end of their thermonuclear evolution. Supernovae are so powerful that they create new atomic nuclei in its interior where the central densities are several times higher than nuclear saturation density \cite{Glendenning:1991es}. Thus, the internal composition and the properties of matter are largely unknown to us. The data available from laboratory experiments strongly indicate the possible existence of quark matter inside NS. The usual model used for QSs is a bag model, but here we assume EoS based on homogeneous and unpaired, overall electrically neutral, 3-flavor interacting quark matter \cite{Flores:2017kte}. Within this theory, one can describe this phase using the simple thermodynamic Bag model EoS \cite{Alford:2004pf} with $\mathcal{O}$ $(m_s^4)$ corrections. With these assumptions, several authors have examined the roles of these interacting parameter $a_4$ and provide relatively tight
constraints on the quark matter EoS, see e.g.~Refs.~\cite{Becerra-Vergara:2019uzm, Banerjee:2020dad, Panotopoulos:2021sbf}. The QCD motivated EoS can be addressed via explicit expressions like \cite{Becerra-Vergara:2019uzm}
\begin{align} \label{Prad1}
  p =&\ \dfrac{1}{3}\left(\rho-4B\right)-\dfrac{m_{s}^{2}}{3\pi}\sqrt{\dfrac{\rho-B}{a_4}}  \nonumber  \\
  &+ \dfrac{m_{s}^{4}}{12\pi^{2}}\left[1-\dfrac{1}{a_4}+3\ln\left(\dfrac{8\pi}{3m_{s}^{2}}\sqrt{\dfrac{\rho-B}{a_4}}\right)\right],
\end{align}
where $\rho$ is the energy density of homogeneously distributed quark matter (also to $\mathcal{O}$ $(m_s^4)$ in the Bag model). The role of $a_4$ was examined in Ref.~\cite{Alford:2004pf} and the authors showed that for $a_4 \approx 0.7$, one can obtain  $M \approx 2 M_{\odot}$ hybrid stars while still ensuring that the nuclear matter to quark matter phase transition occurs above the nuclear saturation density. In that regard, Tangphati \textit{et al.}  \cite{Tangphati:2021tcy, Tangphati:2021mvu} have proposed several QS models and showed direct influence of the interacting parameter on mass-radius relations of QSs.  The Bag constant $B$ lies within the range of  $57\leq B \leq 92\, \rm MeV/fm^3$ \cite{Burgio:2018mcr, Blaschke:2018mqw} and the strange quark mass $m_{s}$ to be $100\, {\rm MeV}$ \cite{Beringer:2012}. In the present work, we will use $B = 60\, \rm MeV/fm^3$ and $a_4 = 0.7$ for the interacting parameter.

In addition, we need to specify a charge density distribution in order to describe charged quark stars. Following a procedure analogous to that carried out in standard GR \cite{Ray:2003gt, Arbanil2013}, we assume that the electric charge density is proportional to the energy density of standard matter, namely
\begin{equation}\label{CDprofile}
  \rho_{\rm ch} = \alpha \rho ,
\end{equation}
where the constant $\alpha$ is a free parameter which measures the amount of charge within the stellar fluid. Here we will adopt values of $\alpha$ for which appreciable changes in the mass-radius diagrams can be observed.


\section{Numerical results}\label{sec4}

Before continuing with the study of charged equilibrium configurations in $f(R,T)$ gravity, we should point out that the stellar structure differential equations in this work are solved numerically using the fourth order Runge-Kutta method in ``Wolfram Mathematica'' software.

\begin{figure*}
    \includegraphics[width=8.5cm]{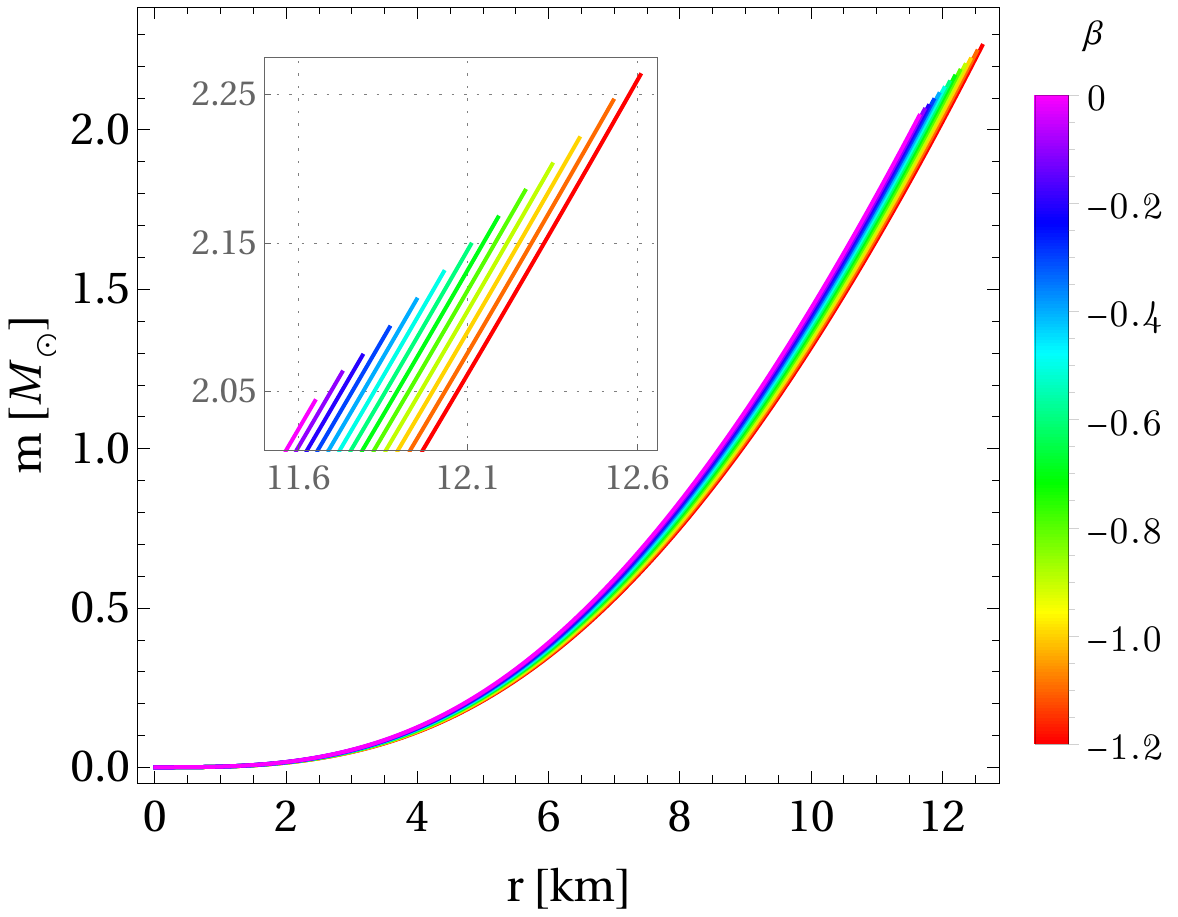}
    \includegraphics[width=8.55cm]{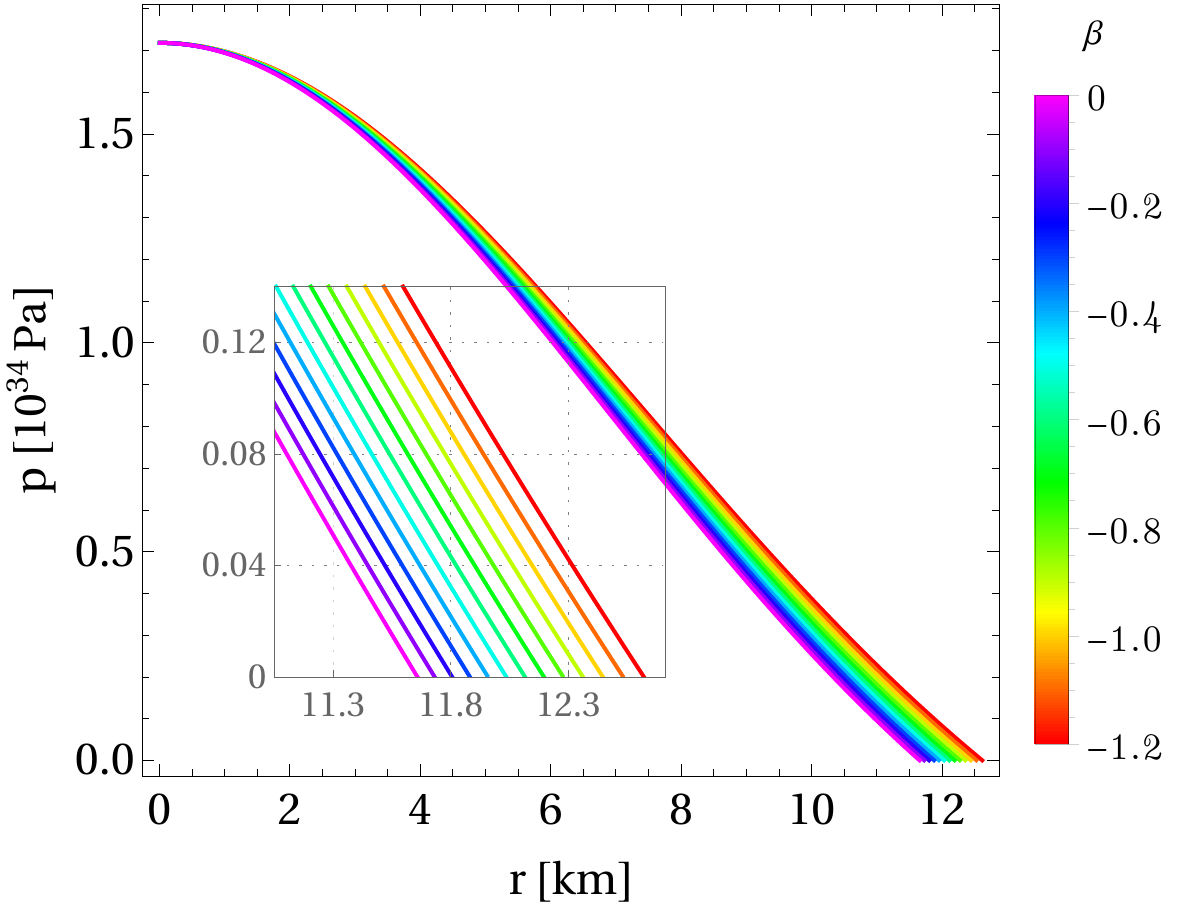}
    \caption{ Numerical solution of the modified TOV equations (\ref{TOV1})-(\ref{TOV3}) with boundary conditions (\ref{BC1}) and EoS (\ref{Prad1}) for a given central energy density $\rho_c = 1.0 \times 10^{18}\rm kg/m^3$, charge parameter $\alpha = 0.4$, and $\beta$ varying in the range $\beta \in [-1.2, 0]$. The radius of the star $r_{\rm sur}$ is determined when the pressure vanishes, and the total mass $M$ is calculated at the surface. Left panel: Mass function, where a smaller value of $\beta$ results in larger masses and radii. Right panel: Pressure as a function of the radial coordinate.  Here we have adopted the set of constants $B= 60 \rm MeV/fm^3$, $a_4 = 0.7$ and $m_s = 100 \rm MeV$.  }
    \label{figure1}
\end{figure*}

\begin{figure*}
    \includegraphics[width=8.5cm]{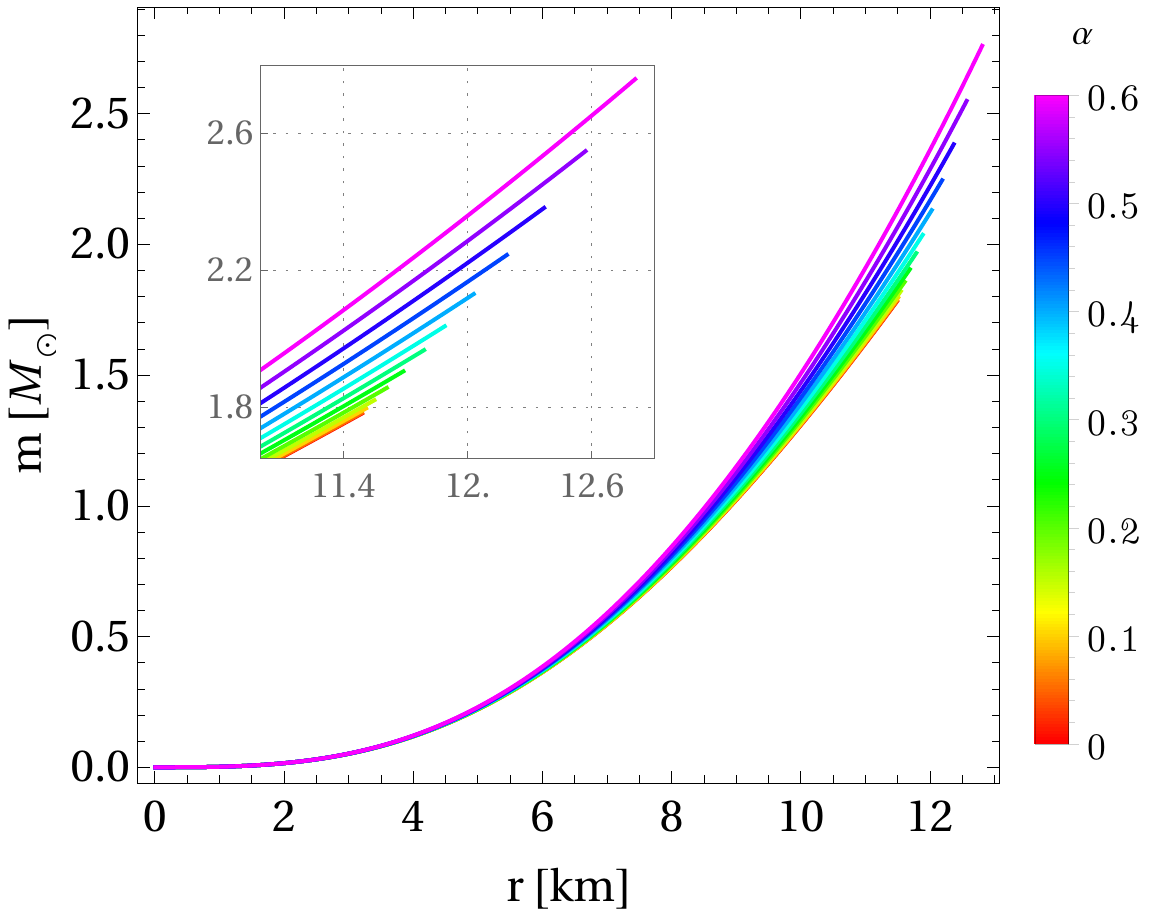}
    \includegraphics[width=8.55cm]{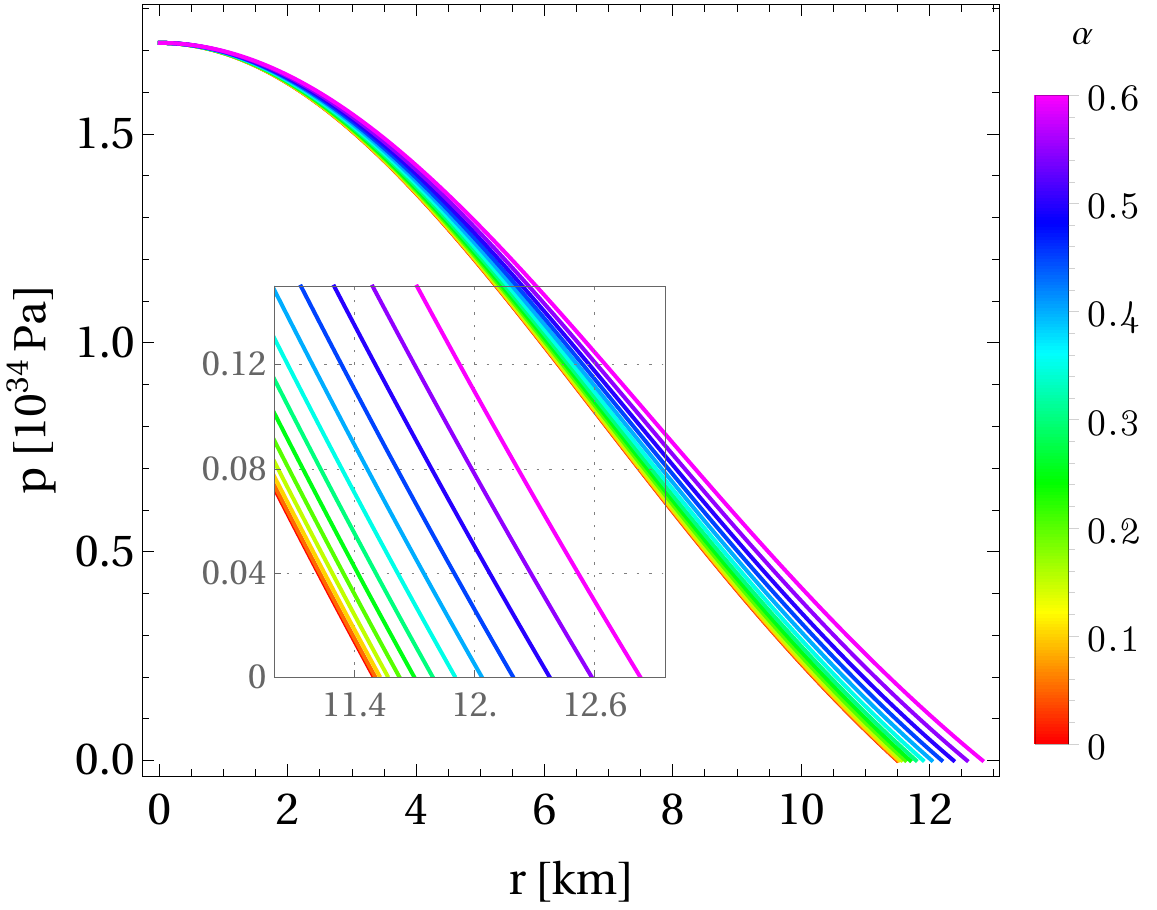}
    \caption{ Radial behavior of the mass function (left panel) and pressure (right panel) for a central energy density $\rho_c = 1.0 \times 10^{18}\rm kg/m^3$, free constant $\beta = -0.5$, and the charge parameter varying in the range $\alpha \in [0, 0.6]$. It can be observed that both the total mass and the radius of the star increase as $\alpha$ increases. The particular case $\alpha =0$ in red corresponds to the uncharged solution. The constants $B$, $a_4$ and $m_s$ assume the same values as in Fig.~\ref{figure1}. }
    \label{figure2}
\end{figure*}

Analogous to the procedure carried out in GR, the system of modified TOV equations (\ref{TOV1})-(\ref{TOV3}) with boundary conditions (\ref{BC1}) will be numerically integrated from the origin at $r= 0$ to the surface of the star at $r= r_{\rm sur}$. In particular, for a central energy density $\rho_c = 1.0 \times 10^{18}\rm kg/m^3$ with EoS (\ref{Prad1}), Fig.~\ref{figure1} displays the mass function and pressure as functions of the radial coordinate for a specific value of $\alpha = 0.4$ and different values of $\beta$. We note that the impact of the $2\beta T$ term on the stellar structure is a significant increase in both the radius and the total mass of the star. Furthermore, if we keep the value of $\beta$ fixed and vary the parameter $\alpha$, it is possible to analyze the effect of the electric charge on a QS within the context of $f(R,T)= R+ 2\beta T$ gravity. According to Fig.~\ref{figure2}, we observe that both the total gravitational mass and the radius of the star increase as the charge parameter increases. 

\begin{figure*}
    \includegraphics[width=8.5cm]{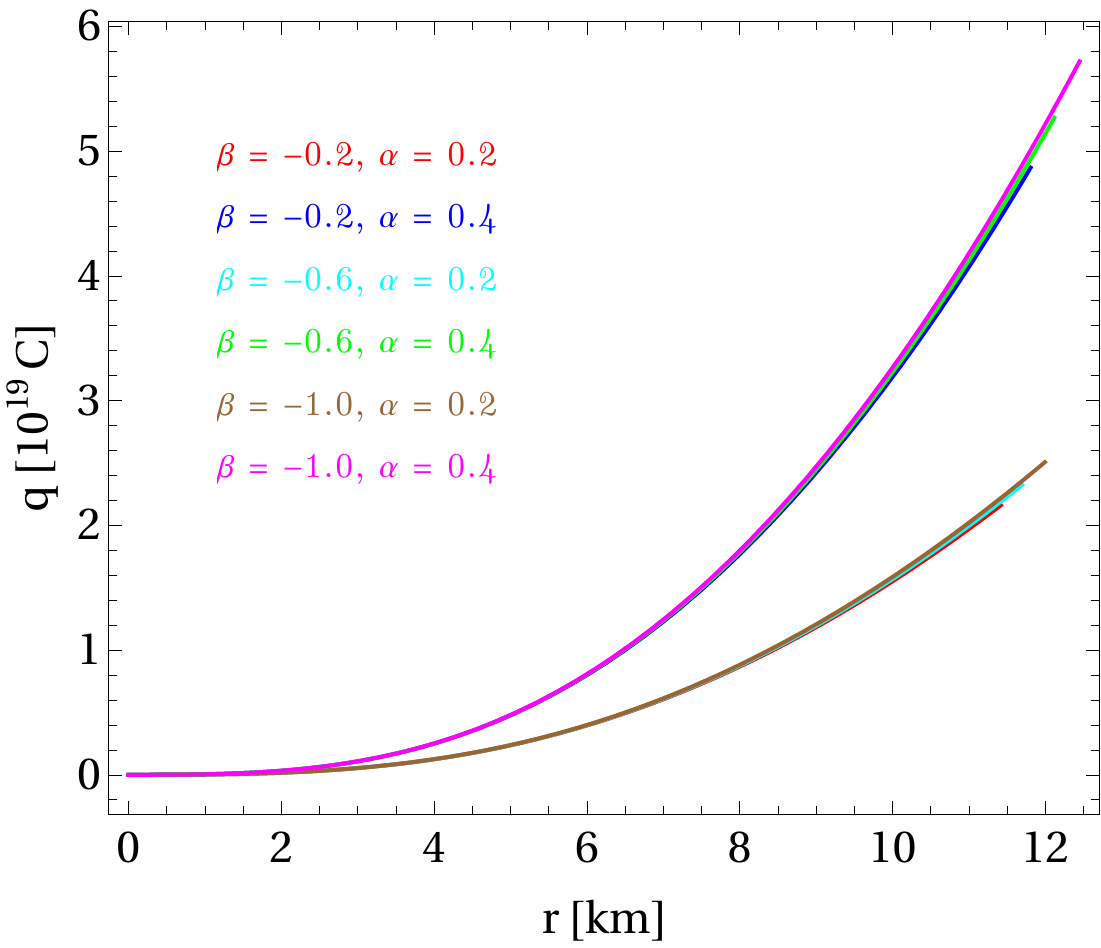}
    \includegraphics[width=8.78cm]{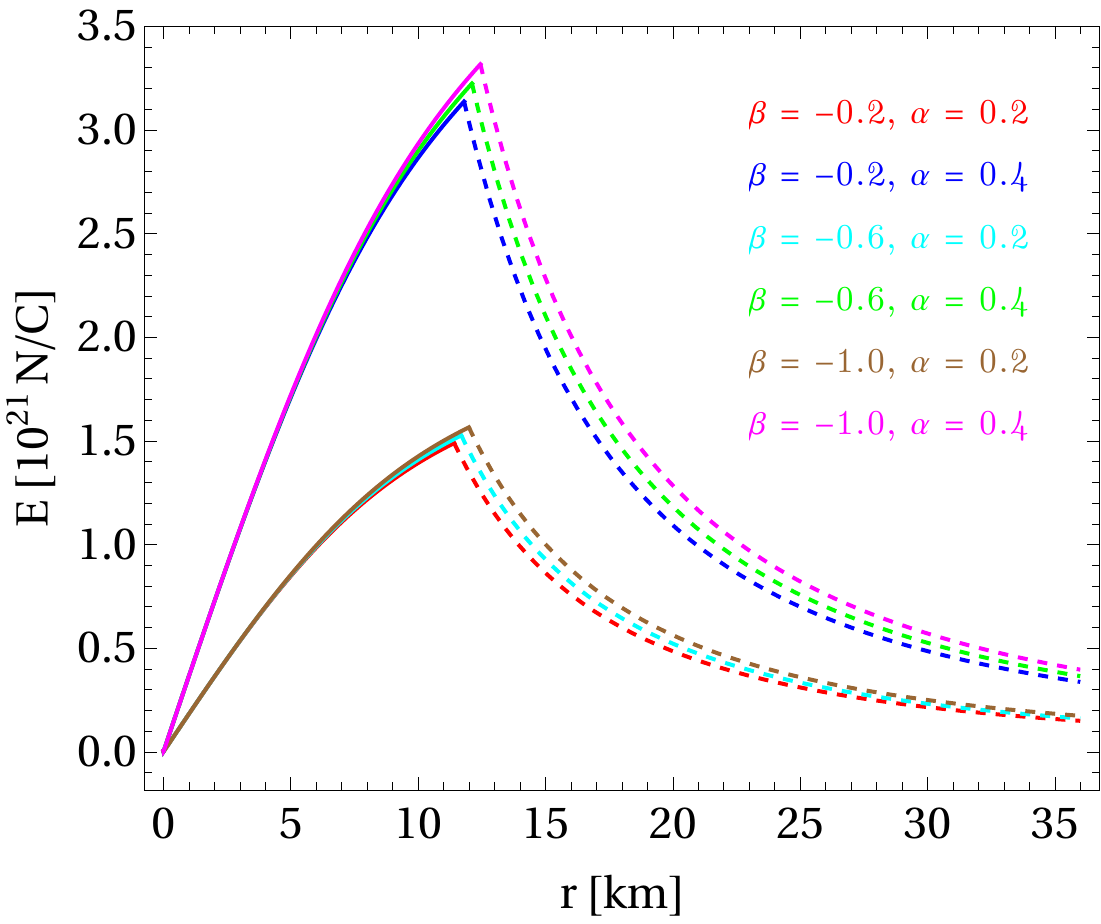}
    \caption{Radial profile of electric charge (left panel) and electric field (right panel) for the central energy density considered in Fig.~\ref{figure1} and several values of $\alpha$ and $\beta$. The dashed curves on the right plot indicate the electric field in the exterior region of the star, which goes to zero as we approach infinity. One can observe that the most relevant changes in both electric charge and field are introduced by the parameter $\alpha$, while such quantities vary slightly as $\beta$ varies and $\alpha$ is fixed. Notice that the main consequence of the parameter $\beta$ is a significant increase in the radius of the star. }
    \label{figure3}
\end{figure*}

We also allow both $\alpha$ and $\beta$ to vary in Fig.~\ref{figure3}, where we show the radial behavior of the charge distribution $q(r)$ and electric field $E(r) = q(r)/4\pi\epsilon_0 r^2$ for several values of $\alpha$ and $\beta$. As expected, the electric charge is an increasing function and $E(r)$ decreases as we move away from the stellar surface. The total charge (determined at the stellar surface) increases as $\beta$ becomes more negative. Moreover, it can be observed that the greatest effects are obtained when we vary the charge parameter $\alpha$, while the main consequence of varying $\beta$ is an increase in the radius of the star.

\begin{figure*}
    \includegraphics[width=8.58cm]{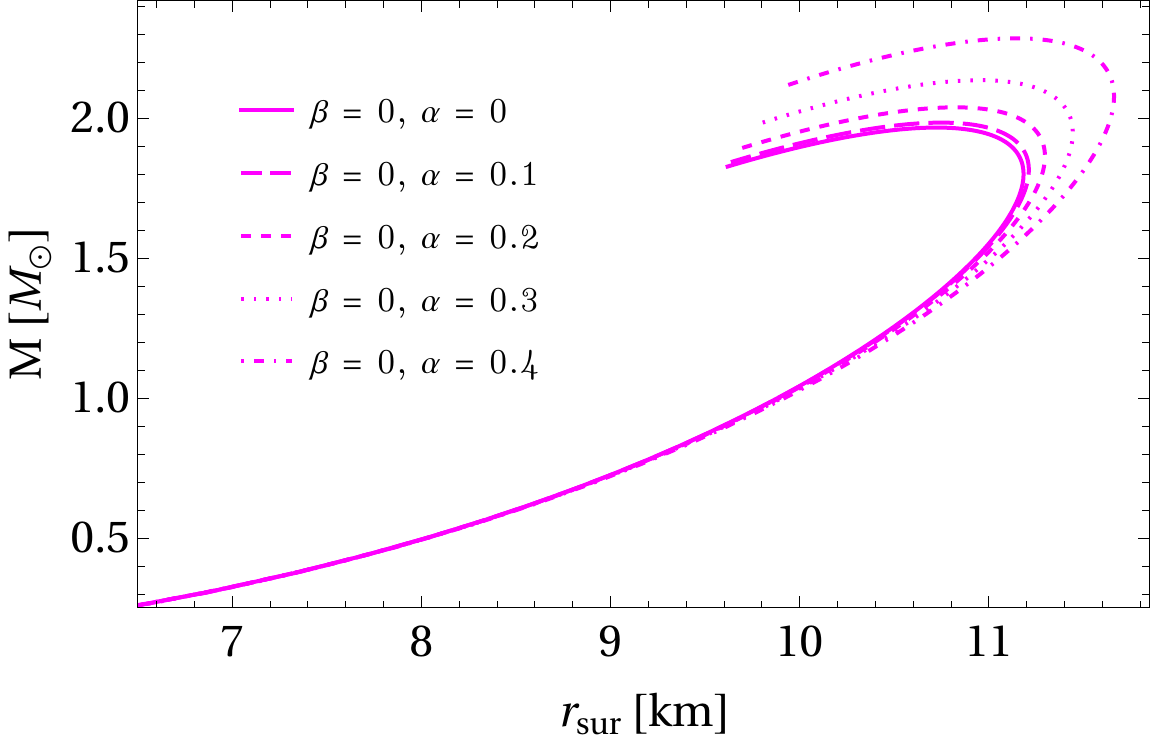}
    \includegraphics[width=8.5cm]{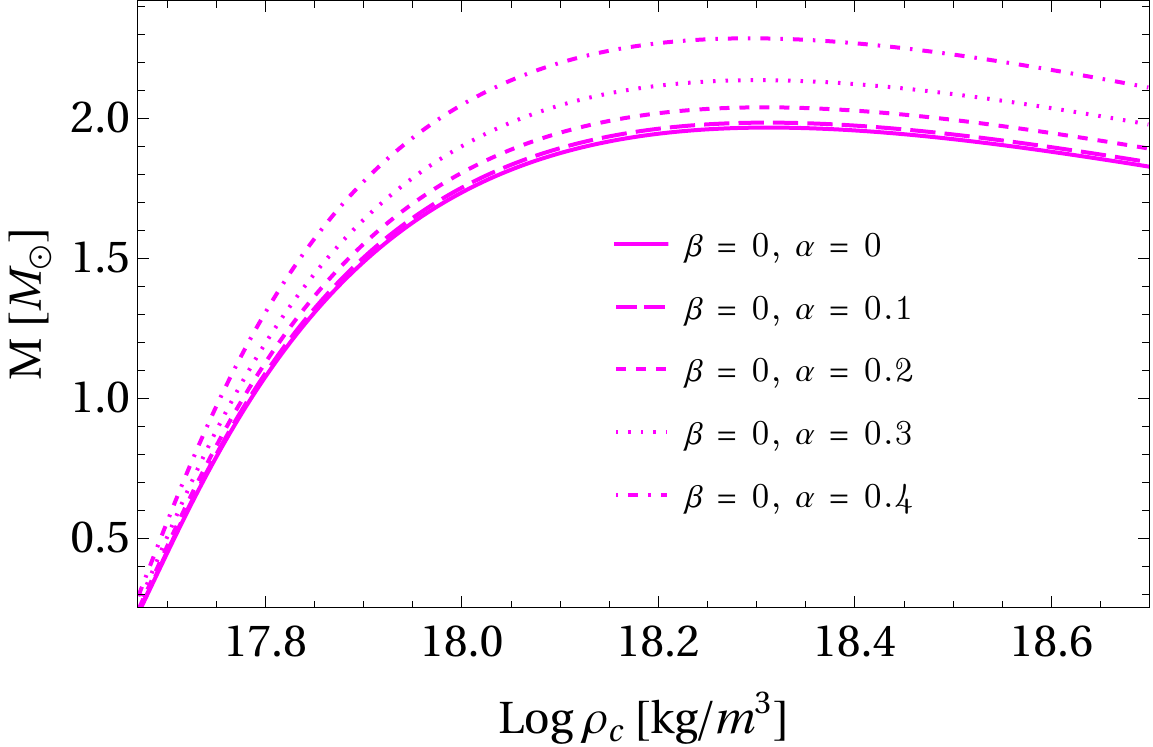}
    \includegraphics[width=8.58cm]{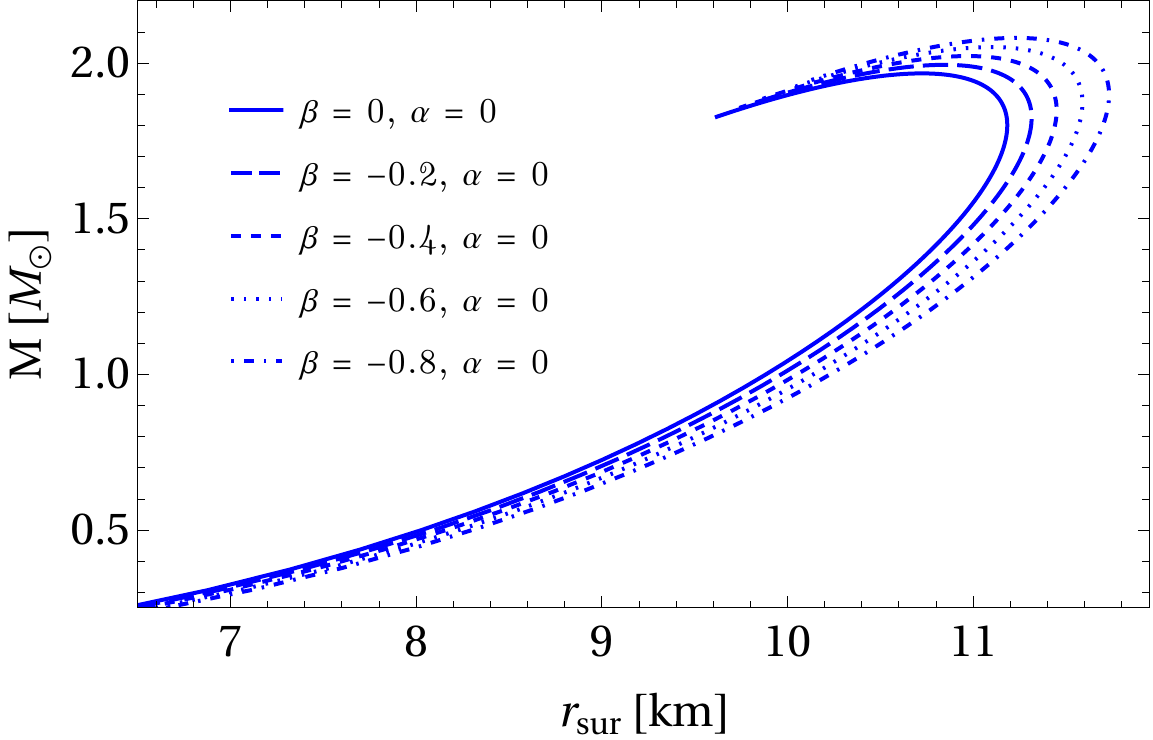}
    \includegraphics[width=8.5cm]{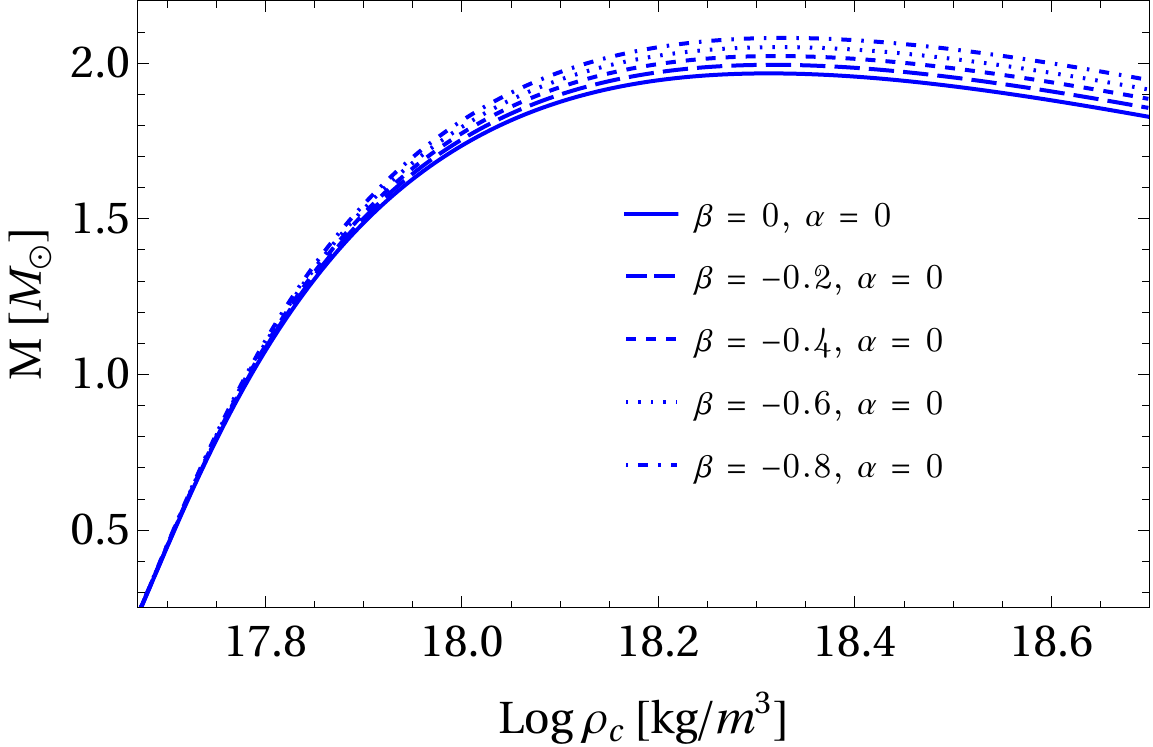}
    \includegraphics[width=8.58cm]{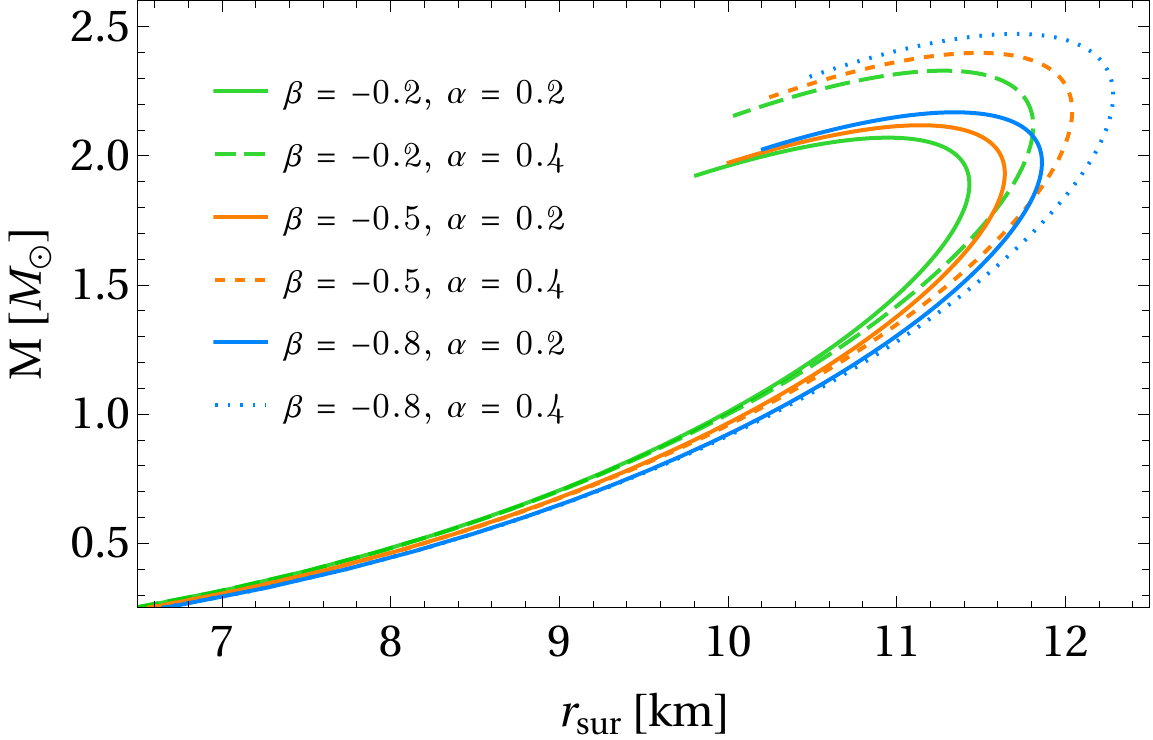}
    \includegraphics[width=8.5cm]{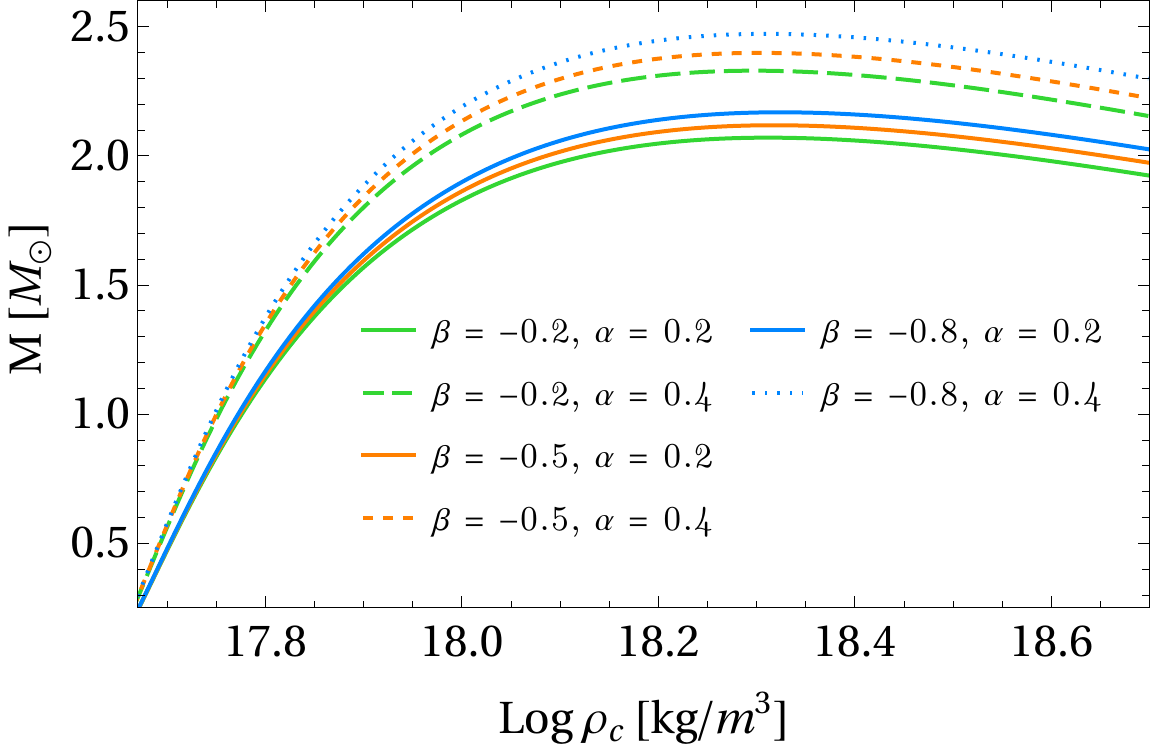}
    \caption{ Mass-radius relations (left column) and mass-central density curves (right column) for quark stars with interacting quark EoS in GR and within the framework of $f(R,T) = R+ 2\beta T$ gravity for different values of $\beta$. The magenta curves in the upper panels represent pure general relativistic solutions for charged stars, i.e. for $\beta= 0$ and $\alpha \neq 0$. Blue lines in the middle panels correspond to uncharged solutions ($\alpha =0$) for different values of $\beta$. The lower plots correspond to $\alpha \neq 0$ and $\beta\neq 0$. Substantial changes due to electric charge occur in the high-mass region. Furthermore, the extra term $2\beta T$ plays a crucial role in significantly modifying the radius of the stars. Note that both parameters $\alpha$ and $\beta$ allow an increase in the maximum-mass values. The constants $B$, $a_4$ and $m_s$ for the EoS (\ref{Prad1}) assume the same values as in Fig.~\ref{figure1}. }
    \label{figure4}
\end{figure*}

By varying the value of the central energy density, we can obtain a family of QSs in hydrostatic equilibrium, as presented in Fig.~\ref{figure4} through the mass-radius diagrams and mass-central density relations. The charged case in standard Einstein gravity (that is, when $\beta= 0$) is shown in the top panels by the magenta curves. The mass and radius undergo substantial changes due to electric charge only in the high-mass region, while the changes are irrelevant for small masses. On the other hand, when $\alpha=0$ and we vary the values of $\beta$, the intermediate plots exhibit significant deviations from the pure GR for uncharged stars. These results have been included for comparison purposes. Nevertheless, the bottom panels show the effect of the electric charge and modification term $2\beta T$ on the macro-physical properties of QSs when both parameters $\alpha$ and $\beta$ are varied simultaneously. It is visualized that such parameters allow maximum-mass values above $2M_\odot$, and hence an appropriate combination of these can lead to masses that fit the observational data.

\begin{figure}
    \includegraphics[width=8.5cm]{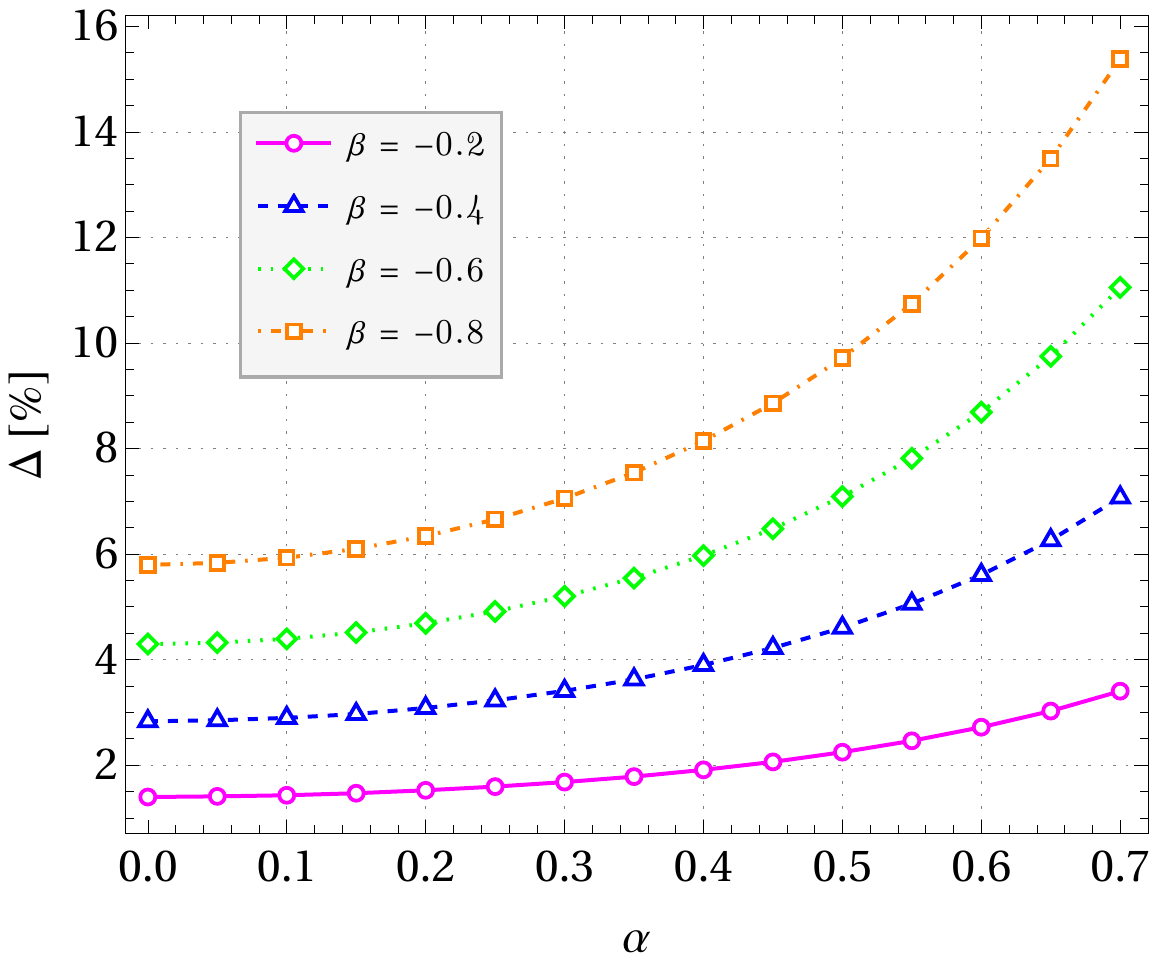}
    \caption{ Relative deviation (\ref{DeltaEq}) as a function of the charge parameter for several fixed values of $\beta$. The uncharged case corresponds to $\alpha =0$, and the deviations grow as $\alpha$ is larger. }
    \label{figure5}
\end{figure}

In order to better quantify the changes in the maximum mass induced by the electric charge within the $f(R,T)= R+ 2\beta T$ gravitational background, we can define a relative difference between the maximum-mass values in GR and $f(R,T)$ gravity, namely
\begin{equation}\label{DeltaEq}
    \Delta = \left. \frac{ M_{{\rm max}, f(R,T)} - M_{\rm max, GR} }{M_{\rm max, GR}} \right\vert_\beta .
\end{equation}
Given a value of $\beta$, such differences can be plotted as a function of the charge parameter, as shown in Fig.~\ref{figure5}. As expected, the deviations are close to zero when $\beta \rightarrow 0$. In the uncharged case ($\alpha= 0$) the deviation grows to about $6\%$ for $\beta= -0.8$. However, with the increase of $\alpha$ the deviation grows, reaching up to $\sim 16\%$ for $\alpha = 0.7$. As a consequence, the largest deviations from GR are achieved for large values of $\alpha$.

\begin{figure*}
    \includegraphics[width=8.5cm]{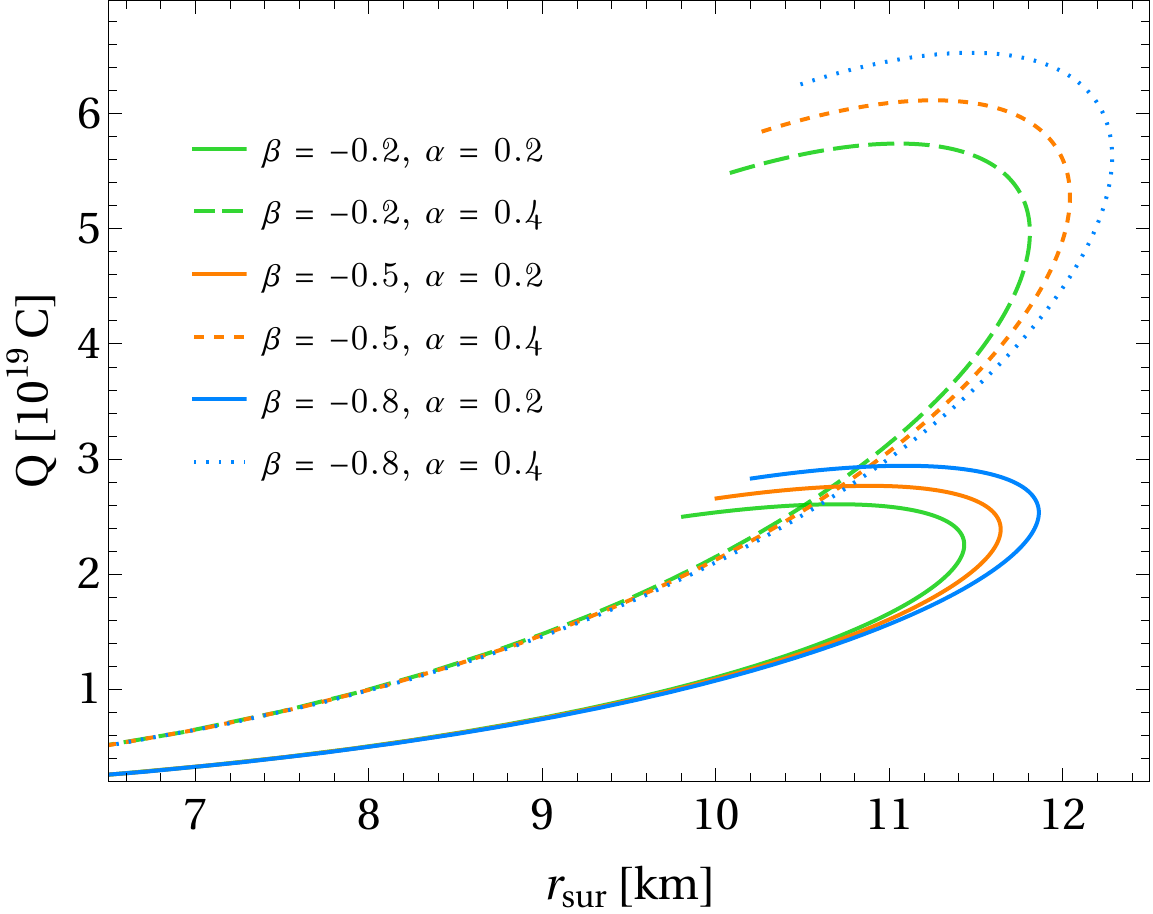}
    \includegraphics[width=8.5cm]{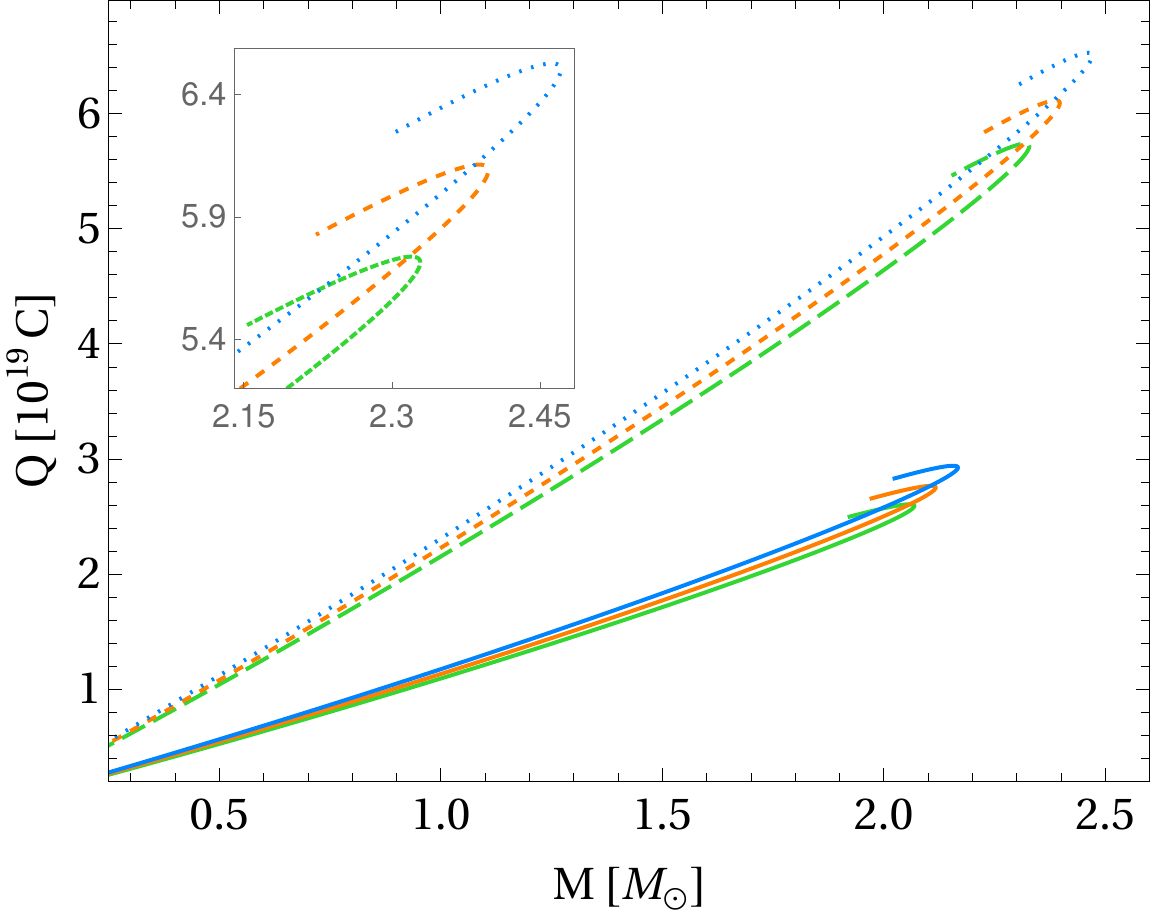}
    \caption{ Left panel: Total charge versus radius predicted by the $f(R,T) = R+ 2\beta T$ gravity model, where different values of $\beta$ and $\alpha$ are considered for the EoS (\ref{Prad1}). Right panel: Total charge plotted as a function of mass, where more negative values of $\beta$ yield higher maximum-charge values. The most relevant modifications due to the parameter $\beta$ occur in the high-charge region. Moreover, as expected, the total charge increases as $\alpha$ increases. }
    \label{figure6}
\end{figure*}

The total charge versus radius is plotted on the left panel of Fig.~\ref{figure6}. Similar to the trends in the pure GR case, larger values of $\alpha$ lead to a larger total charge. Nevertheless, it should be noted that when $\alpha$ assumes a fixed value, the total charge undergoes relevant alterations only in the high-charge region due to the extra term $2\beta T$. Furthermore, the total charge as a function of the gravitational mass is displayed on the right panel of the same figure, where larger masses also yield higher charges but details are less apparent when $\alpha$ is fixed.

Very recently, the authors in Ref.~\cite{Pretel2022} have constructed charged quark stars in pure $f(R)$ gravity for a quadratic function in the Ricci scalar, i.e., the Starobinsky model. Nonetheless, in the present work we have considered a function $f(R,T)$ linear in $R$ but with an additional contribution from the trace of the energy-momentum tensor by means of the $2\beta T$ term. From the mass-radius diagrams (for the uncharged case) we can observe that the radius undergoes relevant changes due to the $2\beta T$ term throughout the curve, while in the Starobinsky model the radius is significantly altered only in the high-mass region.

\section{Concluding remarks }\label{sec5}

In the present paper, we have investigated the physical properties of compact stars made of a charged perfect fluid in the context of modified $f(R,T)$ gravity. An important feature of the theory is an extra force in the geodesic equation of a perfect fluid which may account for the accelerated expansion of the universe. The matter field was assumed to be an interacting EoS (\ref{Prad1}) which strictly depends on the interaction parameter, $a_4$ and the bag constant, $B$.  Moreover, the EoS reduces to the simplest MIT bag model when the quark mass $m_s =0$. In particular, assuming that the charge distribution is proportional to the energy density, we obtained the basic features describing the structure of charged compact objects in $f(R,T)$ gravity. 

The modified TOV equations for this model were solved numerically with appropriate boundary conditions, and subsequently we obtained the structure of spherically symmetric QSs. In the next step, we have studied the mass-radius and mass-central density relations depending on the value of (i) $\beta= 0$ and $\alpha \neq 0$ (Einstein-Maxwell gravity), (ii) $\beta \neq 0$ and $\alpha = 0$ and (iii) $\beta \neq 0$ and $\alpha \neq 0$.  We have discussed further characteristics of the $f(R,T)$ model on theoretical grounds, identifying the significant deviation of $f(R,T)$ gravity from standard Einstein-Maxwell gravity at the high-mass region.

Depending on the parameters of the model, we observed that varying the value of $\alpha$ and $\beta$ both the total gravitational mass and the radius of the star increase. All these features can be seen in Figs.~\ref{figure1} and \ref{figure2}, respectively. In the case $\beta =0$, i.e. in Einstein-Maxwell gravity, the mass and radius undergo substantial changes due to electric charge only in the  high-mass region, while the changes are negligible for small masses. Another interesting feature is the effect of the electric charge and modification term $2\beta T$ on QSs when both parameters $\alpha$ and $\beta$ are varied simultaneously. Our considerations show that considerably increasing the value of $\alpha$ and decreasing the value of $\beta$, a significant increasing of maximum mass ($M_\text{max} > 2M_\odot$) is possible, see bottom panels of Fig.~\ref{figure4}.

The next step was to show the deviation between the maximum-mass values in GR and $f(R,T)$ model. The calculations showed that, when $\alpha= 0$, i.e.~for the uncharged case, the deviation grows to about $6\%$ whereas for $\alpha \neq 0$ the deviation grows up to about $16\%$. This indicates that the presence of larger values of $\alpha$ would lead to more massive QSs from their standard GR counterparts, as we can see from Fig.~\ref{figure5}. Finally, we can say that our formalism in $f(R,T)$ gravity can be applied to analyze the structure of QSs with other dense-matter EoSs.

The advantage of using the particular gravity model $f(R, T)= R+ 2\beta T$ is that it allows us to obtain $R=0$ outside the star, and hence the exterior spacetime can still be described by the Reissner-Nordstr{\"o}m metric. This would not be valid for non-linear functions in $R$ because the scalar curvature does not vanish in the outer region of the star. In fact, the non-zero extra contribution around the star due to such curvature in pure $f(R)$ gravity has been interpreted as a ``gravitational sphere'' in the literature \cite{Astashenok2015, Astashenok2017, Pretel2022CQG}. The price to pay for considering non-linear functions is that a second-order differential equation for the Ricci scalar emerges from the trace of the field equations and, consequently, the complexity of the problem increases. 

On the other hand, we must point out that the model considered in this work has been subject to criticism in recent years, although there is still an open discussion. Fisher and Carlson \cite{Fisher2019} have argued that the term $f_2(T)$ for the separable function $f(R,T)= f_1(R) + f_2(T)$ should be included in the matter Lagrangian $\mathcal{L}_m$ and thus has no physical significance. Nevertheless, Harko and Moraes \cite{Harko2020PRD} showed that there are several major conceptual problems related to the physical interpretations reexamined by the authors in Ref.~\cite{Fisher2019}.

\begin{acknowledgments}
 JMZP acknowledges financial support from the PCI program of the Brazilian agency ``Conselho Nacional de Desenvolvimento Cient{\'i}fico e Tecnol{\'o}gico''--CNPq. T. Tangphati was supported by King Mongkut's University of Technology Thonburi's Post-doctoral Fellowship. A. Pradhan thanks IUCCA, Pune, India for providing facilities under associateship programmes.
\end{acknowledgments}\


%


\newpage
\begin{thebibliography}{70}%
\makeatletter
\providecommand \@ifxundefined [1]{%
 \@ifx{#1\undefined}
}%
\providecommand \@ifnum [1]{%
 \ifnum #1\expandafter \@firstoftwo
 \else \expandafter \@secondoftwo
 \fi
}%
\providecommand \@ifx [1]{%
 \ifx #1\expandafter \@firstoftwo
 \else \expandafter \@secondoftwo
 \fi
}%
\providecommand \natexlab [1]{#1}%
\providecommand \enquote  [1]{``#1''}%
\providecommand \bibnamefont  [1]{#1}%
\providecommand \bibfnamefont [1]{#1}%
\providecommand \citenamefont [1]{#1}%
\providecommand \href@noop [0]{\@secondoftwo}%
\providecommand \href [0]{\begingroup \@sanitize@url \@href}%
\providecommand \@href[1]{\@@startlink{#1}\@@href}%
\providecommand \@@href[1]{\endgroup#1\@@endlink}%
\providecommand \@sanitize@url [0]{\catcode `\\12\catcode `\$12\catcode
  `\&12\catcode `\#12\catcode `\^12\catcode `\_12\catcode `\%12\relax}%
\providecommand \@@startlink[1]{}%
\providecommand \@@endlink[0]{}%
\providecommand \url  [0]{\begingroup\@sanitize@url \@url }%
\providecommand \@url [1]{\endgroup\@href {#1}{\urlprefix }}%
\providecommand \urlprefix  [0]{URL }%
\providecommand \Eprint [0]{\href }%
\providecommand \doibase [0]{http://dx.doi.org/}%
\providecommand \selectlanguage [0]{\@gobble}%
\providecommand \bibinfo  [0]{\@secondoftwo}%
\providecommand \bibfield  [0]{\@secondoftwo}%
\providecommand \translation [1]{[#1]}%
\providecommand \BibitemOpen [0]{}%
\providecommand \bibitemStop [0]{}%
\providecommand \bibitemNoStop [0]{.\EOS\space}%
\providecommand \EOS [0]{\spacefactor3000\relax}%
\providecommand \BibitemShut  [1]{\csname bibitem#1\endcsname}%
\let\auto@bib@innerbib\@empty
\bibitem [{\citenamefont {Coley}(2019)}]{Coley:2018mzr}%
  \BibitemOpen
  \bibfield  {author} {\bibinfo {author} {\bibfnamefont {A.~A.}\ \bibnamefont
  {Coley}},\ }\href {\doibase https://doi.org/10.1007/s10714-019-2559-5}
  {\bibfield  {journal} {\bibinfo  {journal} {Gen. Relativ. Gravit.}\ }\textbf
  {\bibinfo {volume} {51}},\ \bibinfo {pages} {78} (\bibinfo {year}
  {2019})}\BibitemShut {NoStop}%
\bibitem [{\citenamefont {Sotiriou}\ and\ \citenamefont
  {Faraoni}(2010)}]{Sotiriou:2008rp}%
  \BibitemOpen
  \bibfield  {author} {\bibinfo {author} {\bibfnamefont {T.~P.}\ \bibnamefont
  {Sotiriou}}\ and\ \bibinfo {author} {\bibfnamefont {V.}~\bibnamefont
  {Faraoni}},\ }\href {\doibase 10.1103/RevModPhys.82.451} {\bibfield
  {journal} {\bibinfo  {journal} {Rev. Mod. Phys.}\ }\textbf {\bibinfo {volume}
  {82}},\ \bibinfo {pages} {451} (\bibinfo {year} {2010})}\BibitemShut
  {NoStop}%
\bibitem [{\citenamefont {{De Felice}}\ and\ \citenamefont
  {{Tsujikawa}}(2010)}]{DeFelice:2010aj}%
  \BibitemOpen
  \bibfield  {author} {\bibinfo {author} {\bibfnamefont {A.}~\bibnamefont {{De
  Felice}}}\ and\ \bibinfo {author} {\bibfnamefont {S.}~\bibnamefont
  {{Tsujikawa}}},\ }\href {\doibase 10.12942/lrr-2010-3} {\bibfield  {journal}
  {\bibinfo  {journal} {Living Rev. Relativ.}\ }\textbf {\bibinfo {volume}
  {13}},\ \bibinfo {eid} {3} (\bibinfo {year} {2010})}\BibitemShut {NoStop}%
\bibitem [{\citenamefont {Nojiri}\ and\ \citenamefont
  {Odintsov}(2011)}]{Nojiri:2010wj}%
  \BibitemOpen
  \bibfield  {author} {\bibinfo {author} {\bibfnamefont {S.}~\bibnamefont
  {Nojiri}}\ and\ \bibinfo {author} {\bibfnamefont {S.~D.}\ \bibnamefont
  {Odintsov}},\ }\href {\doibase https://doi.org/10.1016/j.physrep.2011.04.001}
  {\bibfield  {journal} {\bibinfo  {journal} {Physics Reports}\ }\textbf
  {\bibinfo {volume} {505}},\ \bibinfo {pages} {59} (\bibinfo {year}
  {2011})}\BibitemShut {NoStop}%
\bibitem [{\citenamefont {Nojiri}\ \emph {et~al.}(2017)\citenamefont {Nojiri},
  \citenamefont {Odintsov},\ and\ \citenamefont {Oikonomou}}]{Nojiri:2017ncd}%
  \BibitemOpen
  \bibfield  {author} {\bibinfo {author} {\bibfnamefont {S.}~\bibnamefont
  {Nojiri}}, \bibinfo {author} {\bibfnamefont {S.}~\bibnamefont {Odintsov}}, \
  and\ \bibinfo {author} {\bibfnamefont {V.}~\bibnamefont {Oikonomou}},\ }\href
  {\doibase https://doi.org/10.1016/j.physrep.2017.06.001} {\bibfield
  {journal} {\bibinfo  {journal} {Physics Reports}\ }\textbf {\bibinfo {volume}
  {692}},\ \bibinfo {pages} {1} (\bibinfo {year} {2017})}\BibitemShut {NoStop}%
\bibitem [{\citenamefont {Capozziello}\ \emph {et~al.}(2015)\citenamefont
  {Capozziello}, \citenamefont {Lobo},\ and\ \citenamefont
  {Mimoso}}]{Capozziello:2014bqa}%
  \BibitemOpen
  \bibfield  {author} {\bibinfo {author} {\bibfnamefont {S.}~\bibnamefont
  {Capozziello}}, \bibinfo {author} {\bibfnamefont {F.~S.~N.}\ \bibnamefont
  {Lobo}}, \ and\ \bibinfo {author} {\bibfnamefont {J.~P.}\ \bibnamefont
  {Mimoso}},\ }\href {\doibase 10.1103/PhysRevD.91.124019} {\bibfield
  {journal} {\bibinfo  {journal} {Phys. Rev. D}\ }\textbf {\bibinfo {volume}
  {91}},\ \bibinfo {pages} {124019} (\bibinfo {year} {2015})}\BibitemShut
  {NoStop}%
\bibitem [{\citenamefont {Bertolami}\ \emph {et~al.}(2010)\citenamefont
  {Bertolami}, \citenamefont {Fraz\~ao},\ and\ \citenamefont
  {P\'aramos}}]{Bertolami:2010cw}%
  \BibitemOpen
  \bibfield  {author} {\bibinfo {author} {\bibfnamefont {O.}~\bibnamefont
  {Bertolami}}, \bibinfo {author} {\bibfnamefont {P.}~\bibnamefont {Fraz\~ao}},
  \ and\ \bibinfo {author} {\bibfnamefont {J.}~\bibnamefont {P\'aramos}},\
  }\href {\doibase 10.1103/PhysRevD.81.104046} {\bibfield  {journal} {\bibinfo
  {journal} {Phys. Rev. D}\ }\textbf {\bibinfo {volume} {81}},\ \bibinfo
  {pages} {104046} (\bibinfo {year} {2010})}\BibitemShut {NoStop}%
\bibitem [{\citenamefont {Harko}\ \emph {et~al.}(2011)\citenamefont {Harko},
  \citenamefont {Lobo}, \citenamefont {Nojiri},\ and\ \citenamefont
  {Odintsov}}]{Harko2011}%
  \BibitemOpen
  \bibfield  {author} {\bibinfo {author} {\bibfnamefont {T.}~\bibnamefont
  {Harko}}, \bibinfo {author} {\bibfnamefont {F.~S.~N.}\ \bibnamefont {Lobo}},
  \bibinfo {author} {\bibfnamefont {S.}~\bibnamefont {Nojiri}}, \ and\ \bibinfo
  {author} {\bibfnamefont {S.~D.}\ \bibnamefont {Odintsov}},\ }\href {\doibase
  10.1103/PhysRevD.84.024020} {\bibfield  {journal} {\bibinfo  {journal} {Phys.
  Rev. D}\ }\textbf {\bibinfo {volume} {84}},\ \bibinfo {pages} {024020}
  (\bibinfo {year} {2011})}\BibitemShut {NoStop}%
\bibitem [{\citenamefont {Shabani}\ and\ \citenamefont
  {Farhoudi}(2013)}]{Shabani:2013djy}%
  \BibitemOpen
  \bibfield  {author} {\bibinfo {author} {\bibfnamefont {H.}~\bibnamefont
  {Shabani}}\ and\ \bibinfo {author} {\bibfnamefont {M.}~\bibnamefont
  {Farhoudi}},\ }\href {\doibase 10.1103/PhysRevD.88.044048} {\bibfield
  {journal} {\bibinfo  {journal} {Phys. Rev. D}\ }\textbf {\bibinfo {volume}
  {88}},\ \bibinfo {pages} {044048} (\bibinfo {year} {2013})}\BibitemShut
  {NoStop}%
\bibitem [{\citenamefont {Alvarenga}\ \emph {et~al.}(2013)\citenamefont
  {Alvarenga}, \citenamefont {Houndjo}, \citenamefont {Monwanou},\ and\
  \citenamefont {Orou}}]{Alvarenga:2012bt}%
  \BibitemOpen
  \bibfield  {author} {\bibinfo {author} {\bibfnamefont {F.~G.}\ \bibnamefont
  {Alvarenga}}, \bibinfo {author} {\bibfnamefont {M.~J.~S.}\ \bibnamefont
  {Houndjo}}, \bibinfo {author} {\bibfnamefont {A.~V.}\ \bibnamefont
  {Monwanou}}, \ and\ \bibinfo {author} {\bibfnamefont {J.~B.~C.}\ \bibnamefont
  {Orou}},\ }\href {\doibase https://doi.org/10.4236/jmp.2013.41019} {\bibfield
   {journal} {\bibinfo  {journal} {J. Mod. Phys.}\ }\textbf {\bibinfo {volume}
  {4}},\ \bibinfo {pages} {130} (\bibinfo {year} {2013})}\BibitemShut {NoStop}%
\bibitem [{\citenamefont {Sharif}\ and\ \citenamefont
  {Zubair}(2013{\natexlab{a}})}]{Sharif:2012ce}%
  \BibitemOpen
  \bibfield  {author} {\bibinfo {author} {\bibfnamefont {M.}~\bibnamefont
  {Sharif}}\ and\ \bibinfo {author} {\bibfnamefont {M.}~\bibnamefont
  {Zubair}},\ }\href {\doibase https://doi.org/10.7566/JPSJ.82.014002}
  {\bibfield  {journal} {\bibinfo  {journal} {J. Phys. Soc. Jap.}\ }\textbf
  {\bibinfo {volume} {82}},\ \bibinfo {pages} {014002} (\bibinfo {year}
  {2013}{\natexlab{a}})}\BibitemShut {NoStop}%
\bibitem [{\citenamefont {Sharif}\ and\ \citenamefont
  {Zubair}(2012)}]{Sharif:2012zzd}%
  \BibitemOpen
  \bibfield  {author} {\bibinfo {author} {\bibfnamefont {M.}~\bibnamefont
  {Sharif}}\ and\ \bibinfo {author} {\bibfnamefont {M.}~\bibnamefont
  {Zubair}},\ }\href {\doibase https://doi.org/10.1088/1475-7516/2012/03/028}
  {\bibfield  {journal} {\bibinfo  {journal} {JCAP}\ }\textbf {\bibinfo
  {volume} {03}},\ \bibinfo {pages} {028} (\bibinfo {year} {2012})}\BibitemShut
  {NoStop}%
\bibitem [{\citenamefont {Jamil}\ \emph {et~al.}(2012)\citenamefont {Jamil},
  \citenamefont {Momeni},\ and\ \citenamefont {Myrzakulov}}]{Jamil:2012pf}%
  \BibitemOpen
  \bibfield  {author} {\bibinfo {author} {\bibfnamefont {M.}~\bibnamefont
  {Jamil}}, \bibinfo {author} {\bibfnamefont {D.}~\bibnamefont {Momeni}}, \
  and\ \bibinfo {author} {\bibfnamefont {R.}~\bibnamefont {Myrzakulov}},\
  }\href {\doibase https://doi.org/10.1088/0256-307X/29/10/109801} {\bibfield
  {journal} {\bibinfo  {journal} {Chinese Phys. Lett.}\ }\textbf {\bibinfo
  {volume} {29}},\ \bibinfo {pages} {109801} (\bibinfo {year}
  {2012})}\BibitemShut {NoStop}%
\bibitem [{\citenamefont {Sharif}\ and\ \citenamefont
  {Zubair}(2013{\natexlab{b}})}]{Sharif:2013ffa}%
  \BibitemOpen
  \bibfield  {author} {\bibinfo {author} {\bibfnamefont {M.}~\bibnamefont
  {Sharif}}\ and\ \bibinfo {author} {\bibfnamefont {M.}~\bibnamefont
  {Zubair}},\ }\href {\doibase https://doi.org/10.1134/S1063776113100075}
  {\bibfield  {journal} {\bibinfo  {journal} {J. Exp. Theor. Phys.}\ }\textbf
  {\bibinfo {volume} {117}},\ \bibinfo {pages} {248} (\bibinfo {year}
  {2013}{\natexlab{b}})}\BibitemShut {NoStop}%
\bibitem [{\citenamefont {Shabani}\ and\ \citenamefont
  {Farhoudi}(2014)}]{Shabani:2014xvi}%
  \BibitemOpen
  \bibfield  {author} {\bibinfo {author} {\bibfnamefont {H.}~\bibnamefont
  {Shabani}}\ and\ \bibinfo {author} {\bibfnamefont {M.}~\bibnamefont
  {Farhoudi}},\ }\href {\doibase 10.1103/PhysRevD.90.044031} {\bibfield
  {journal} {\bibinfo  {journal} {Phys. Rev. D}\ }\textbf {\bibinfo {volume}
  {90}},\ \bibinfo {pages} {044031} (\bibinfo {year} {2014})}\BibitemShut
  {NoStop}%
\bibitem [{\citenamefont {Xu}\ \emph {et~al.}(2016)\citenamefont {Xu},
  \citenamefont {Harko},\ and\ \citenamefont {Liang}}]{Xu:2016rdf}%
  \BibitemOpen
  \bibfield  {author} {\bibinfo {author} {\bibfnamefont {M.~X.}\ \bibnamefont
  {Xu}}, \bibinfo {author} {\bibfnamefont {T.}~\bibnamefont {Harko}}, \ and\
  \bibinfo {author} {\bibfnamefont {S.~D.}\ \bibnamefont {Liang}},\ }\href
  {\doibase https://doi.org/10.1140/epjc/s10052-016-4303-6} {\bibfield
  {journal} {\bibinfo  {journal} {Eur. Phys. J. C}\ }\textbf {\bibinfo {volume}
  {76}},\ \bibinfo {pages} {449} (\bibinfo {year} {2016})}\BibitemShut
  {NoStop}%
\bibitem [{\citenamefont {Velten}\ and\ \citenamefont
  {Caram\^es}(2017)}]{Velten2017}%
  \BibitemOpen
  \bibfield  {author} {\bibinfo {author} {\bibfnamefont {H.}~\bibnamefont
  {Velten}}\ and\ \bibinfo {author} {\bibfnamefont {T.~R.~P.}\ \bibnamefont
  {Caram\^es}},\ }\href {\doibase 10.1103/PhysRevD.95.123536} {\bibfield
  {journal} {\bibinfo  {journal} {Phys. Rev. D}\ }\textbf {\bibinfo {volume}
  {95}},\ \bibinfo {pages} {123536} (\bibinfo {year} {2017})}\BibitemShut
  {NoStop}%
\bibitem [{\citenamefont {Godani}(2019)}]{Godani2019}%
  \BibitemOpen
  \bibfield  {author} {\bibinfo {author} {\bibfnamefont {N.}~\bibnamefont
  {Godani}},\ }\href {\doibase https://doi.org/10.1142/S0219887819500245}
  {\bibfield  {journal} {\bibinfo  {journal} {Int. J. Geom. Meth. Mod. Phys.}\
  }\textbf {\bibinfo {volume} {16}},\ \bibinfo {pages} {1950024} (\bibinfo
  {year} {2019})}\BibitemShut {NoStop}%
\bibitem [{\citenamefont {Gamonal}(2021)}]{Gamonal2020}%
  \BibitemOpen
  \bibfield  {author} {\bibinfo {author} {\bibfnamefont {M.}~\bibnamefont
  {Gamonal}},\ }\href {\doibase https://doi.org/10.1016/j.dark.2020.100768}
  {\bibfield  {journal} {\bibinfo  {journal} {Phys. Dark Universe}\ }\textbf
  {\bibinfo {volume} {31}},\ \bibinfo {pages} {100768} (\bibinfo {year}
  {2021})}\BibitemShut {NoStop}%
\bibitem [{\citenamefont {Moraes}\ \emph {et~al.}(2016)\citenamefont {Moraes},
  \citenamefont {Arba{\~n}il},\ and\ \citenamefont {Malheiro}}]{Moraes2016}%
  \BibitemOpen
  \bibfield  {author} {\bibinfo {author} {\bibfnamefont {P.~H.~R.~S.}\
  \bibnamefont {Moraes}}, \bibinfo {author} {\bibfnamefont {J.~D.~V.}\
  \bibnamefont {Arba{\~n}il}}, \ and\ \bibinfo {author} {\bibfnamefont
  {M.}~\bibnamefont {Malheiro}},\ }\href {\doibase
  https://doi.org/10.1088/1475-7516/2016/06/005} {\bibfield  {journal}
  {\bibinfo  {journal} {JCAP}\ }\textbf {\bibinfo {volume} {06}},\ \bibinfo
  {pages} {005} (\bibinfo {year} {2016})}\BibitemShut {NoStop}%
\bibitem [{\citenamefont {Das}\ \emph {et~al.}(2016)\citenamefont {Das},
  \citenamefont {Rahaman}, \citenamefont {Guha},\ and\ \citenamefont
  {Ray}}]{Das:2016mxq}%
  \BibitemOpen
  \bibfield  {author} {\bibinfo {author} {\bibfnamefont {A.}~\bibnamefont
  {Das}}, \bibinfo {author} {\bibfnamefont {F.}~\bibnamefont {Rahaman}},
  \bibinfo {author} {\bibfnamefont {B.~K.}\ \bibnamefont {Guha}}, \ and\
  \bibinfo {author} {\bibfnamefont {S.}~\bibnamefont {Ray}},\ }\href {\doibase
  https://doi.org/10.1140/epjc/s10052-016-4503-0} {\bibfield  {journal}
  {\bibinfo  {journal} {Eur. Phys. J. C}\ }\textbf {\bibinfo {volume} {76}},\
  \bibinfo {pages} {654} (\bibinfo {year} {2016})}\BibitemShut {NoStop}%
\bibitem [{\citenamefont {Deb}\ \emph {et~al.}(2018{\natexlab{a}})\citenamefont
  {Deb}, \citenamefont {Rahaman}, \citenamefont {Ray},\ and\ \citenamefont
  {Guha}}]{Deb:2017rhd}%
  \BibitemOpen
  \bibfield  {author} {\bibinfo {author} {\bibfnamefont {D.}~\bibnamefont
  {Deb}}, \bibinfo {author} {\bibfnamefont {F.}~\bibnamefont {Rahaman}},
  \bibinfo {author} {\bibfnamefont {S.}~\bibnamefont {Ray}}, \ and\ \bibinfo
  {author} {\bibfnamefont {B.~K.}\ \bibnamefont {Guha}},\ }\href {\doibase
  https://doi.org/10.1088/1475-7516/2018/03/044} {\bibfield  {journal}
  {\bibinfo  {journal} {JCAP}\ }\textbf {\bibinfo {volume} {03}},\ \bibinfo
  {pages} {044} (\bibinfo {year} {2018}{\natexlab{a}})}\BibitemShut {NoStop}%
\bibitem [{\citenamefont {Biswas}\ \emph {et~al.}(2019)\citenamefont {Biswas}
  \emph {et~al.}}]{Biswas:2018inc}%
  \BibitemOpen
  \bibfield  {author} {\bibinfo {author} {\bibfnamefont {S.}~\bibnamefont
  {Biswas}} \emph {et~al.},\ }\href {\doibase
  https://doi.org/10.1016/j.aop.2018.12.004} {\bibfield  {journal} {\bibinfo
  {journal} {Annals Phys.}\ }\textbf {\bibinfo {volume} {401}},\ \bibinfo
  {pages} {1} (\bibinfo {year} {2019})}\BibitemShut {NoStop}%
\bibitem [{\citenamefont {Lobato}\ \emph {et~al.}(2020)\citenamefont {Lobato}
  \emph {et~al.}}]{Lobato2020}%
  \BibitemOpen
  \bibfield  {author} {\bibinfo {author} {\bibfnamefont {R.}~\bibnamefont
  {Lobato}} \emph {et~al.},\ }\href {\doibase 10.1088/1475-7516/2020/12/039}
  {\bibfield  {journal} {\bibinfo  {journal} {JCAP}\ }\textbf {\bibinfo
  {volume} {12}},\ \bibinfo {pages} {039} (\bibinfo {year} {2020})}\BibitemShut
  {NoStop}%
\bibitem [{\citenamefont {Pretel}\ \emph
  {et~al.}(2021{\natexlab{a}})\citenamefont {Pretel}, \citenamefont
  {Jor{\'{a}}s}, \citenamefont {Reis},\ and\ \citenamefont
  {Arba{\~{n}}il}}]{Pretel2021}%
  \BibitemOpen
  \bibfield  {author} {\bibinfo {author} {\bibfnamefont {J.~M.~Z.}\
  \bibnamefont {Pretel}}, \bibinfo {author} {\bibfnamefont {S.~E.}\
  \bibnamefont {Jor{\'{a}}s}}, \bibinfo {author} {\bibfnamefont {R.~R.~R.}\
  \bibnamefont {Reis}}, \ and\ \bibinfo {author} {\bibfnamefont {J.~D.~V.}\
  \bibnamefont {Arba{\~{n}}il}},\ }\href {\doibase
  10.1088/1475-7516/2021/08/055} {\bibfield  {journal} {\bibinfo  {journal}
  {JCAP}\ }\textbf {\bibinfo {volume} {08}},\ \bibinfo {pages} {055} (\bibinfo
  {year} {2021}{\natexlab{a}})}\BibitemShut {NoStop}%
\bibitem [{\citenamefont {Pretel}\ \emph
  {et~al.}(2021{\natexlab{b}})\citenamefont {Pretel}, \citenamefont
  {Jor{\'a}s}, \citenamefont {Reis},\ and\ \citenamefont
  {Arba{\~n}il}}]{Pretel:2020oae}%
  \BibitemOpen
  \bibfield  {author} {\bibinfo {author} {\bibfnamefont {J.~M.~Z.}\
  \bibnamefont {Pretel}}, \bibinfo {author} {\bibfnamefont {S.~E.}\
  \bibnamefont {Jor{\'a}s}}, \bibinfo {author} {\bibfnamefont {R.~R.~R.}\
  \bibnamefont {Reis}}, \ and\ \bibinfo {author} {\bibfnamefont {J.~D.~V.}\
  \bibnamefont {Arba{\~n}il}},\ }\href {\doibase 10.1088/1475-7516/2021/04/064}
  {\bibfield  {journal} {\bibinfo  {journal} {JCAP}\ }\textbf {\bibinfo
  {volume} {04}},\ \bibinfo {pages} {064} (\bibinfo {year}
  {2021}{\natexlab{b}})}\BibitemShut {NoStop}%
\bibitem [{\citenamefont {Hansraj}(2018)}]{Hansraj:2018cmj}%
  \BibitemOpen
  \bibfield  {author} {\bibinfo {author} {\bibfnamefont {S.}~\bibnamefont
  {Hansraj}},\ }\href {\doibase https://doi.org/10.1140/epjc/s10052-018-6194-1}
  {\bibfield  {journal} {\bibinfo  {journal} {Eur. Phys. J. C}\ }\textbf
  {\bibinfo {volume} {78}},\ \bibinfo {pages} {700} (\bibinfo {year}
  {2018})}\BibitemShut {NoStop}%
\bibitem [{\citenamefont {Maurya}\ \emph {et~al.}(2020)\citenamefont {Maurya},
  \citenamefont {Banerjee},\ and\ \citenamefont
  {Tello-Ortiz}}]{Maurya:2019msr}%
  \BibitemOpen
  \bibfield  {author} {\bibinfo {author} {\bibfnamefont {S.~K.}\ \bibnamefont
  {Maurya}}, \bibinfo {author} {\bibfnamefont {A.}~\bibnamefont {Banerjee}}, \
  and\ \bibinfo {author} {\bibfnamefont {F.}~\bibnamefont {Tello-Ortiz}},\
  }\href {\doibase https://doi.org/10.1016/j.dark.2019.100438} {\bibfield
  {journal} {\bibinfo  {journal} {Phys. Dark Univ.}\ }\textbf {\bibinfo
  {volume} {27}},\ \bibinfo {pages} {100438} (\bibinfo {year}
  {2020})}\BibitemShut {NoStop}%
\bibitem [{\citenamefont {Deb}\ \emph {et~al.}(2018{\natexlab{b}})\citenamefont
  {Deb}, \citenamefont {Guha}, \citenamefont {Rahaman},\ and\ \citenamefont
  {Ray}}]{Deb:2017rhc}%
  \BibitemOpen
  \bibfield  {author} {\bibinfo {author} {\bibfnamefont {D.}~\bibnamefont
  {Deb}}, \bibinfo {author} {\bibfnamefont {B.~K.}\ \bibnamefont {Guha}},
  \bibinfo {author} {\bibfnamefont {F.}~\bibnamefont {Rahaman}}, \ and\
  \bibinfo {author} {\bibfnamefont {S.}~\bibnamefont {Ray}},\ }\href {\doibase
  10.1103/PhysRevD.97.084026} {\bibfield  {journal} {\bibinfo  {journal} {Phys.
  Rev. D}\ }\textbf {\bibinfo {volume} {97}},\ \bibinfo {pages} {084026}
  (\bibinfo {year} {2018}{\natexlab{b}})}\BibitemShut {NoStop}%
\bibitem [{\citenamefont {Deb}\ \emph {et~al.}(2019{\natexlab{a}})\citenamefont
  {Deb} \emph {et~al.}}]{Deb:2018sgt}%
  \BibitemOpen
  \bibfield  {author} {\bibinfo {author} {\bibfnamefont {D.}~\bibnamefont
  {Deb}} \emph {et~al.},\ }\href {\doibase
  https://doi.org/10.1093/mnras/stz708} {\bibfield  {journal} {\bibinfo
  {journal} {MNRAS}\ }\textbf {\bibinfo {volume} {485}},\ \bibinfo {pages}
  {5652} (\bibinfo {year} {2019}{\natexlab{a}})}\BibitemShut {NoStop}%
\bibitem [{\citenamefont {Maurya}\ \emph {et~al.}(2019)\citenamefont {Maurya}
  \emph {et~al.}}]{Maurya:2019sfm}%
  \BibitemOpen
  \bibfield  {author} {\bibinfo {author} {\bibfnamefont {S.~K.}\ \bibnamefont
  {Maurya}} \emph {et~al.},\ }\href {\doibase 10.1103/PhysRevD.100.044014}
  {\bibfield  {journal} {\bibinfo  {journal} {Phys. Rev. D}\ }\textbf {\bibinfo
  {volume} {100}},\ \bibinfo {pages} {044014} (\bibinfo {year}
  {2019})}\BibitemShut {NoStop}%
\bibitem [{\citenamefont {Yousaf}\ \emph {et~al.}(2018)\citenamefont {Yousaf},
  \citenamefont {u.~H.~Bhatti},\ and\ \citenamefont {Ilyas}}]{Yousaf:2018jkb}%
  \BibitemOpen
  \bibfield  {author} {\bibinfo {author} {\bibfnamefont {Z.}~\bibnamefont
  {Yousaf}}, \bibinfo {author} {\bibfnamefont {M.~Z.}\ \bibnamefont
  {u.~H.~Bhatti}}, \ and\ \bibinfo {author} {\bibfnamefont {M.}~\bibnamefont
  {Ilyas}},\ }\href {\doibase https://doi.org/10.1140/epjc/s10052-018-5797-x}
  {\bibfield  {journal} {\bibinfo  {journal} {Eur. Phys. J. C}\ }\textbf
  {\bibinfo {volume} {78}},\ \bibinfo {pages} {307} (\bibinfo {year}
  {2018})}\BibitemShut {NoStop}%
\bibitem [{\citenamefont {Hansraj}\ and\ \citenamefont
  {Banerjee}(2018)}]{Hansraj:2018jzb}%
  \BibitemOpen
  \bibfield  {author} {\bibinfo {author} {\bibfnamefont {S.}~\bibnamefont
  {Hansraj}}\ and\ \bibinfo {author} {\bibfnamefont {A.}~\bibnamefont
  {Banerjee}},\ }\href {\doibase 10.1103/PhysRevD.97.104020} {\bibfield
  {journal} {\bibinfo  {journal} {Phys. Rev. D}\ }\textbf {\bibinfo {volume}
  {97}},\ \bibinfo {pages} {104020} (\bibinfo {year} {2018})}\BibitemShut
  {NoStop}%
\bibitem [{\citenamefont {Das}\ \emph {et~al.}(2017)\citenamefont {Das} \emph
  {et~al.}}]{Das:2017rhi}%
  \BibitemOpen
  \bibfield  {author} {\bibinfo {author} {\bibfnamefont {A.}~\bibnamefont
  {Das}} \emph {et~al.},\ }\href {\doibase 10.1103/PhysRevD.95.124011}
  {\bibfield  {journal} {\bibinfo  {journal} {Phys. Rev. D}\ }\textbf {\bibinfo
  {volume} {95}},\ \bibinfo {pages} {124011} (\bibinfo {year}
  {2017})}\BibitemShut {NoStop}%
\bibitem [{\citenamefont {Moraes}\ and\ \citenamefont
  {Sahoo}(2017)}]{Moraes:2017mir}%
  \BibitemOpen
  \bibfield  {author} {\bibinfo {author} {\bibfnamefont {P.~H. R.~S.}\
  \bibnamefont {Moraes}}\ and\ \bibinfo {author} {\bibfnamefont {P.~K.}\
  \bibnamefont {Sahoo}},\ }\href {\doibase 10.1103/PhysRevD.96.044038}
  {\bibfield  {journal} {\bibinfo  {journal} {Phys. Rev. D}\ }\textbf {\bibinfo
  {volume} {96}},\ \bibinfo {pages} {044038} (\bibinfo {year}
  {2017})}\BibitemShut {NoStop}%
\bibitem [{\citenamefont {Banerjee}\ \emph {et~al.}(2020)\citenamefont
  {Banerjee}, \citenamefont {Singh}, \citenamefont {Jasim},\ and\ \citenamefont
  {Rahaman}}]{Banerjee:2019wjj}%
  \BibitemOpen
  \bibfield  {author} {\bibinfo {author} {\bibfnamefont {A.}~\bibnamefont
  {Banerjee}}, \bibinfo {author} {\bibfnamefont {K.~N.}\ \bibnamefont {Singh}},
  \bibinfo {author} {\bibfnamefont {M.~K.}\ \bibnamefont {Jasim}}, \ and\
  \bibinfo {author} {\bibfnamefont {F.}~\bibnamefont {Rahaman}},\ }\href
  {\doibase https://doi.org/10.1016/j.aop.2020.168295} {\bibfield  {journal}
  {\bibinfo  {journal} {Annals Phys.}\ }\textbf {\bibinfo {volume} {422}},\
  \bibinfo {pages} {168295} (\bibinfo {year} {2020})}\BibitemShut {NoStop}%
\bibitem [{\citenamefont {Elizalde}\ and\ \citenamefont
  {Khurshudyan}(2019)}]{Elizalde:2018arz}%
  \BibitemOpen
  \bibfield  {author} {\bibinfo {author} {\bibfnamefont {E.}~\bibnamefont
  {Elizalde}}\ and\ \bibinfo {author} {\bibfnamefont {M.}~\bibnamefont
  {Khurshudyan}},\ }\href {\doibase 10.1103/PhysRevD.99.024051} {\bibfield
  {journal} {\bibinfo  {journal} {Phys. Rev. D}\ }\textbf {\bibinfo {volume}
  {99}},\ \bibinfo {pages} {024051} (\bibinfo {year} {2019})}\BibitemShut
  {NoStop}%
\bibitem [{\citenamefont {Deb}\ \emph {et~al.}(2019{\natexlab{b}})\citenamefont
  {Deb}, \citenamefont {Ketov}, \citenamefont {Khlopov},\ and\ \citenamefont
  {Ray}}]{Deb:2018gzt}%
  \BibitemOpen
  \bibfield  {author} {\bibinfo {author} {\bibfnamefont {D.}~\bibnamefont
  {Deb}}, \bibinfo {author} {\bibfnamefont {S.~V.}\ \bibnamefont {Ketov}},
  \bibinfo {author} {\bibfnamefont {M.}~\bibnamefont {Khlopov}}, \ and\
  \bibinfo {author} {\bibfnamefont {S.}~\bibnamefont {Ray}},\ }\href {\doibase
  https://doi.org/10.1088/1475-7516/2019/10/070} {\bibfield  {journal}
  {\bibinfo  {journal} {JCAP}\ }\textbf {\bibinfo {volume} {10}},\ \bibinfo
  {pages} {070} (\bibinfo {year} {2019}{\natexlab{b}})}\BibitemShut {NoStop}%
\bibitem [{\citenamefont {Sharif}\ and\ \citenamefont
  {Waseem}(2018)}]{Sharif:2018khl}%
  \BibitemOpen
  \bibfield  {author} {\bibinfo {author} {\bibfnamefont {M.}~\bibnamefont
  {Sharif}}\ and\ \bibinfo {author} {\bibfnamefont {A.}~\bibnamefont
  {Waseem}},\ }\href {\doibase https://doi.org/10.1140/epjc/s10052-018-6363-2}
  {\bibfield  {journal} {\bibinfo  {journal} {Eur. Phys. J. C}\ }\textbf
  {\bibinfo {volume} {78}},\ \bibinfo {pages} {868} (\bibinfo {year}
  {2018})}\BibitemShut {NoStop}%
\bibitem [{\citenamefont {Tangphati}\ \emph
  {et~al.}(2022{\natexlab{a}})\citenamefont {Tangphati}, \citenamefont
  {Hansraj}, \citenamefont {Banerjee},\ and\ \citenamefont
  {Pradhan}}]{Tangphati:2022mur}%
  \BibitemOpen
  \bibfield  {author} {\bibinfo {author} {\bibfnamefont {T.}~\bibnamefont
  {Tangphati}}, \bibinfo {author} {\bibfnamefont {S.}~\bibnamefont {Hansraj}},
  \bibinfo {author} {\bibfnamefont {A.}~\bibnamefont {Banerjee}}, \ and\
  \bibinfo {author} {\bibfnamefont {A.}~\bibnamefont {Pradhan}},\ }\href
  {\doibase https://doi.org/10.1016/j.dark.2022.100990} {\bibfield  {journal}
  {\bibinfo  {journal} {Phys. Dark Universe}\ }\textbf {\bibinfo {volume}
  {35}},\ \bibinfo {pages} {100990} (\bibinfo {year}
  {2022}{\natexlab{a}})}\BibitemShut {NoStop}%
\bibitem [{\citenamefont {Tangphati}\ \emph
  {et~al.}(2022{\natexlab{b}})\citenamefont {Tangphati}, \citenamefont {Karar},
  \citenamefont {Pradhan},\ and\ \citenamefont {Banerjee}}]{Tangphati:2022arm}%
  \BibitemOpen
  \bibfield  {author} {\bibinfo {author} {\bibfnamefont {T.}~\bibnamefont
  {Tangphati}}, \bibinfo {author} {\bibfnamefont {I.}~\bibnamefont {Karar}},
  \bibinfo {author} {\bibfnamefont {A.}~\bibnamefont {Pradhan}}, \ and\
  \bibinfo {author} {\bibfnamefont {A.}~\bibnamefont {Banerjee}},\ }\href
  {\doibase https://doi.org/10.1140/epjc/s10052-022-10024-6} {\bibfield
  {journal} {\bibinfo  {journal} {Eur. Phys. J. C}\ }\textbf {\bibinfo {volume}
  {82}},\ \bibinfo {pages} {57} (\bibinfo {year}
  {2022}{\natexlab{b}})}\BibitemShut {NoStop}%
\bibitem [{\citenamefont {Eddington}(1988)}]{Eddington}%
  \BibitemOpen
  \bibfield  {author} {\bibinfo {author} {\bibfnamefont {A.~S.}\ \bibnamefont
  {Eddington}},\ }\href@noop {} {\emph {\bibinfo {title} {Internal Constitution
  of the Stars}}}\ (\bibinfo  {publisher} {Cambridge University Press},\
  \bibinfo {year} {1988})\BibitemShut {NoStop}%
\bibitem [{\citenamefont {Olson}\ and\ \citenamefont {Bailyn}(1975)}]{Olson}%
  \BibitemOpen
  \bibfield  {author} {\bibinfo {author} {\bibfnamefont {E.}~\bibnamefont
  {Olson}}\ and\ \bibinfo {author} {\bibfnamefont {M.}~\bibnamefont {Bailyn}},\
  }\href {\doibase 10.1103/PhysRevD.12.3030} {\bibfield  {journal} {\bibinfo
  {journal} {Phys. Rev. D}\ }\textbf {\bibinfo {volume} {12}},\ \bibinfo
  {pages} {3030} (\bibinfo {year} {1975})}\BibitemShut {NoStop}%
\bibitem [{\citenamefont {Olson}\ and\ \citenamefont {Bailyn}(1976)}]{Olson1}%
  \BibitemOpen
  \bibfield  {author} {\bibinfo {author} {\bibfnamefont {E.}~\bibnamefont
  {Olson}}\ and\ \bibinfo {author} {\bibfnamefont {M.}~\bibnamefont {Bailyn}},\
  }\href {\doibase 10.1103/PhysRevD.13.2204} {\bibfield  {journal} {\bibinfo
  {journal} {Phys. Rev. D}\ }\textbf {\bibinfo {volume} {13}},\ \bibinfo
  {pages} {2204} (\bibinfo {year} {1976})}\BibitemShut {NoStop}%
\bibitem [{\citenamefont {{Shvartsman}}(1971)}]{Shvartsman}%
  \BibitemOpen
  \bibfield  {author} {\bibinfo {author} {\bibfnamefont {V.~F.}\ \bibnamefont
  {{Shvartsman}}},\ }\href
  {https://ui.adsabs.harvard.edu/abs/1971JETP...33..475S} {\bibfield  {journal}
  {\bibinfo  {journal} {Soviet Journal of Experimental and Theoretical
  Physics}\ }\textbf {\bibinfo {volume} {33}},\ \bibinfo {pages} {475}
  (\bibinfo {year} {1971})}\BibitemShut {NoStop}%
\bibitem [{\citenamefont {{Alcock}}\ \emph {et~al.}(1986)\citenamefont
  {{Alcock}}, \citenamefont {{Farhi}},\ and\ \citenamefont
  {{Olinto}}}]{alcock86:a}%
  \BibitemOpen
  \bibfield  {author} {\bibinfo {author} {\bibfnamefont {C.}~\bibnamefont
  {{Alcock}}}, \bibinfo {author} {\bibfnamefont {E.}~\bibnamefont {{Farhi}}}, \
  and\ \bibinfo {author} {\bibfnamefont {A.}~\bibnamefont {{Olinto}}},\ }\href
  {\doibase 10.1086/164679} {\bibfield  {journal} {\bibinfo  {journal} {\apj}\
  }\textbf {\bibinfo {volume} {310}},\ \bibinfo {pages} {261} (\bibinfo {year}
  {1986})}\BibitemShut {NoStop}%
\bibitem [{\citenamefont {Alcock}\ and\ \citenamefont
  {Olinto}(1988)}]{alcock88:a}%
  \BibitemOpen
  \bibfield  {author} {\bibinfo {author} {\bibfnamefont {C.}~\bibnamefont
  {Alcock}}\ and\ \bibinfo {author} {\bibfnamefont {A.~V.}\ \bibnamefont
  {Olinto}},\ }\href {\doibase 10.1146/annurev.ns.38.120188.001113} {\bibfield
  {journal} {\bibinfo  {journal} {Ann. Rev. Nucl. Part. Sci.}\ }\textbf
  {\bibinfo {volume} {38}},\ \bibinfo {pages} {161} (\bibinfo {year}
  {1988})}\BibitemShut {NoStop}%
\bibitem [{\citenamefont {Ray}\ \emph {et~al.}(2003)\citenamefont {Ray} \emph
  {et~al.}}]{Ray:2003gt}%
  \BibitemOpen
  \bibfield  {author} {\bibinfo {author} {\bibfnamefont {S.}~\bibnamefont
  {Ray}} \emph {et~al.},\ }\href {\doibase 10.1103/PhysRevD.68.084004}
  {\bibfield  {journal} {\bibinfo  {journal} {Phys. Rev. D}\ }\textbf {\bibinfo
  {volume} {68}},\ \bibinfo {pages} {084004} (\bibinfo {year}
  {2003})}\BibitemShut {NoStop}%
\bibitem [{\citenamefont {Panotopoulos}\ \emph {et~al.}(2022)\citenamefont
  {Panotopoulos}, \citenamefont {Tangphati},\ and\ \citenamefont
  {Banerjee}}]{Panotopoulos:2021cxu}%
  \BibitemOpen
  \bibfield  {author} {\bibinfo {author} {\bibfnamefont {G.}~\bibnamefont
  {Panotopoulos}}, \bibinfo {author} {\bibfnamefont {T.}~\bibnamefont
  {Tangphati}}, \ and\ \bibinfo {author} {\bibfnamefont {A.}~\bibnamefont
  {Banerjee}},\ }\href {\doibase https://doi.org/10.1016/j.cjph.2021.10.027}
  {\bibfield  {journal} {\bibinfo  {journal} {Chin. J. Phys.}\ }\textbf
  {\bibinfo {volume} {77}},\ \bibinfo {pages} {1682} (\bibinfo {year}
  {2022})}\BibitemShut {NoStop}%
\bibitem [{\citenamefont {Lemos}\ \emph {et~al.}(2015)\citenamefont {Lemos},
  \citenamefont {Lopes}, \citenamefont {Quinta},\ and\ \citenamefont
  {Zanchin}}]{Lemos:2014lza}%
  \BibitemOpen
  \bibfield  {author} {\bibinfo {author} {\bibfnamefont {J.~P.~S.}\
  \bibnamefont {Lemos}}, \bibinfo {author} {\bibfnamefont {F.~J.}\ \bibnamefont
  {Lopes}}, \bibinfo {author} {\bibfnamefont {G.}~\bibnamefont {Quinta}}, \
  and\ \bibinfo {author} {\bibfnamefont {V.~T.}\ \bibnamefont {Zanchin}},\
  }\href {\doibase https://doi.org/10.1140/epjc/s10052-015-3274-3} {\bibfield
  {journal} {\bibinfo  {journal} {Eur. Phys. J. C}\ }\textbf {\bibinfo {volume}
  {75}},\ \bibinfo {pages} {76} (\bibinfo {year} {2015})}\BibitemShut {NoStop}%
\bibitem [{\citenamefont {Arba\~nil}\ and\ \citenamefont
  {Zanchin}(2018)}]{Arbanil:2017huq}%
  \BibitemOpen
  \bibfield  {author} {\bibinfo {author} {\bibfnamefont {J.~D.~V.}\
  \bibnamefont {Arba\~nil}}\ and\ \bibinfo {author} {\bibfnamefont {V.~T.}\
  \bibnamefont {Zanchin}},\ }\href {\doibase 10.1103/PhysRevD.97.104045}
  {\bibfield  {journal} {\bibinfo  {journal} {Phys. Rev. D}\ }\textbf {\bibinfo
  {volume} {97}},\ \bibinfo {pages} {104045} (\bibinfo {year}
  {2018})}\BibitemShut {NoStop}%
\bibitem [{\citenamefont {Negreiros}\ \emph {et~al.}(2009)\citenamefont
  {Negreiros}, \citenamefont {Weber}, \citenamefont {Malheiro},\ and\
  \citenamefont {Usov}}]{Negreiros2009}%
  \BibitemOpen
  \bibfield  {author} {\bibinfo {author} {\bibfnamefont {R.~P.}\ \bibnamefont
  {Negreiros}}, \bibinfo {author} {\bibfnamefont {F.}~\bibnamefont {Weber}},
  \bibinfo {author} {\bibfnamefont {M.}~\bibnamefont {Malheiro}}, \ and\
  \bibinfo {author} {\bibfnamefont {V.}~\bibnamefont {Usov}},\ }\href {\doibase
  10.1103/PhysRevD.80.083006} {\bibfield  {journal} {\bibinfo  {journal} {Phys.
  Rev. D}\ }\textbf {\bibinfo {volume} {80}},\ \bibinfo {pages} {083006}
  (\bibinfo {year} {2009})}\BibitemShut {NoStop}%
\bibitem [{\citenamefont {Arba\~nil}\ \emph {et~al.}(2013)\citenamefont
  {Arba\~nil}, \citenamefont {Lemos},\ and\ \citenamefont
  {Zanchin}}]{Arbanil2013}%
  \BibitemOpen
  \bibfield  {author} {\bibinfo {author} {\bibfnamefont {J.~D.~V.}\
  \bibnamefont {Arba\~nil}}, \bibinfo {author} {\bibfnamefont {J.~P.~S.}\
  \bibnamefont {Lemos}}, \ and\ \bibinfo {author} {\bibfnamefont {V.~T.}\
  \bibnamefont {Zanchin}},\ }\href {\doibase 10.1103/PhysRevD.88.084023}
  {\bibfield  {journal} {\bibinfo  {journal} {Phys. Rev. D}\ }\textbf {\bibinfo
  {volume} {88}},\ \bibinfo {pages} {084023} (\bibinfo {year}
  {2013})}\BibitemShut {NoStop}%
\bibitem [{\citenamefont {Pretel}\ \emph {et~al.}(2022)\citenamefont {Pretel}
  \emph {et~al.}}]{Pretel2022}%
  \BibitemOpen
  \bibfield  {author} {\bibinfo {author} {\bibfnamefont {J.~M.~Z.}\
  \bibnamefont {Pretel}} \emph {et~al.},\ }\href
  {https://doi.org/10.48550/arXiv.2206.03878} {\bibfield  {journal} {\bibinfo
  {journal} {arXiv:2206.03878 [gr-qc]}\ } (\bibinfo {year} {2022})}\BibitemShut
  {NoStop}%
\bibitem [{\citenamefont {Glendenning}\ and\ \citenamefont
  {Moszkowski}(1991)}]{Glendenning:1991es}%
  \BibitemOpen
  \bibfield  {author} {\bibinfo {author} {\bibfnamefont {N.~K.}\ \bibnamefont
  {Glendenning}}\ and\ \bibinfo {author} {\bibfnamefont {S.~A.}\ \bibnamefont
  {Moszkowski}},\ }\href {\doibase 10.1103/PhysRevLett.67.2414} {\bibfield
  {journal} {\bibinfo  {journal} {Phys. Rev. Lett.}\ }\textbf {\bibinfo
  {volume} {67}},\ \bibinfo {pages} {2414} (\bibinfo {year}
  {1991})}\BibitemShut {NoStop}%
\bibitem [{\citenamefont {Flores}\ \emph {et~al.}(2017)\citenamefont {Flores},
  \citenamefont {Hall},\ and\ \citenamefont {Jaikumar}}]{Flores:2017kte}%
  \BibitemOpen
  \bibfield  {author} {\bibinfo {author} {\bibfnamefont {C.~V.}\ \bibnamefont
  {Flores}}, \bibinfo {author} {\bibfnamefont {Z.~B.}\ \bibnamefont {Hall}}, \
  and\ \bibinfo {author} {\bibfnamefont {P.}~\bibnamefont {Jaikumar}},\ }\href
  {\doibase 10.1103/PhysRevC.96.065803} {\bibfield  {journal} {\bibinfo
  {journal} {Phys. Rev. C}\ }\textbf {\bibinfo {volume} {96}},\ \bibinfo
  {pages} {065803} (\bibinfo {year} {2017})}\BibitemShut {NoStop}%
\bibitem [{\citenamefont {Alford}\ \emph {et~al.}(2005)\citenamefont {Alford},
  \citenamefont {Braby}, \citenamefont {Paris},\ and\ \citenamefont
  {Reddy}}]{Alford:2004pf}%
  \BibitemOpen
  \bibfield  {author} {\bibinfo {author} {\bibfnamefont {M.}~\bibnamefont
  {Alford}}, \bibinfo {author} {\bibfnamefont {M.}~\bibnamefont {Braby}},
  \bibinfo {author} {\bibfnamefont {M.}~\bibnamefont {Paris}}, \ and\ \bibinfo
  {author} {\bibfnamefont {S.}~\bibnamefont {Reddy}},\ }\href {\doibase
  10.1086/430902} {\bibfield  {journal} {\bibinfo  {journal} {Astrophys. J.}\
  }\textbf {\bibinfo {volume} {629}},\ \bibinfo {pages} {969} (\bibinfo {year}
  {2005})}\BibitemShut {NoStop}%
\bibitem [{\citenamefont {Becerra-Vergara}\ \emph {et~al.}(2019)\citenamefont
  {Becerra-Vergara}, \citenamefont {Mojica}, \citenamefont {Lora-Clavijo},\
  and\ \citenamefont {Cruz-Osorio}}]{Becerra-Vergara:2019uzm}%
  \BibitemOpen
  \bibfield  {author} {\bibinfo {author} {\bibfnamefont {E.~A.}\ \bibnamefont
  {Becerra-Vergara}}, \bibinfo {author} {\bibfnamefont {S.}~\bibnamefont
  {Mojica}}, \bibinfo {author} {\bibfnamefont {F.~D.}\ \bibnamefont
  {Lora-Clavijo}}, \ and\ \bibinfo {author} {\bibfnamefont {A.}~\bibnamefont
  {Cruz-Osorio}},\ }\href {\doibase 10.1103/PhysRevD.100.103006} {\bibfield
  {journal} {\bibinfo  {journal} {Phys. Rev. D}\ }\textbf {\bibinfo {volume}
  {100}},\ \bibinfo {pages} {103006} (\bibinfo {year} {2019})}\BibitemShut
  {NoStop}%
\bibitem [{\citenamefont {Banerjee}\ \emph {et~al.}(2021)\citenamefont
  {Banerjee}, \citenamefont {Tangphati}, \citenamefont {Samart},\ and\
  \citenamefont {Channuie}}]{Banerjee:2020dad}%
  \BibitemOpen
  \bibfield  {author} {\bibinfo {author} {\bibfnamefont {A.}~\bibnamefont
  {Banerjee}}, \bibinfo {author} {\bibfnamefont {T.}~\bibnamefont {Tangphati}},
  \bibinfo {author} {\bibfnamefont {D.}~\bibnamefont {Samart}}, \ and\ \bibinfo
  {author} {\bibfnamefont {P.}~\bibnamefont {Channuie}},\ }\href {\doibase
  10.3847/1538-4357/abc87f} {\bibfield  {journal} {\bibinfo  {journal}
  {Astrophys. J.}\ }\textbf {\bibinfo {volume} {906}},\ \bibinfo {pages} {114}
  (\bibinfo {year} {2021})}\BibitemShut {NoStop}%
\bibitem [{\citenamefont {Panotopoulos}\ \emph {et~al.}(2021)\citenamefont
  {Panotopoulos}, \citenamefont {Tangphati}, \citenamefont {Banerjee},\ and\
  \citenamefont {Jasim}}]{Panotopoulos:2021sbf}%
  \BibitemOpen
  \bibfield  {author} {\bibinfo {author} {\bibfnamefont {G.}~\bibnamefont
  {Panotopoulos}}, \bibinfo {author} {\bibfnamefont {T.}~\bibnamefont
  {Tangphati}}, \bibinfo {author} {\bibfnamefont {A.}~\bibnamefont {Banerjee}},
  \ and\ \bibinfo {author} {\bibfnamefont {M.~K.}\ \bibnamefont {Jasim}},\
  }\href {\doibase https://doi.org/10.1016/j.physletb.2021.136330} {\bibfield
  {journal} {\bibinfo  {journal} {Phys. Lett. B}\ }\textbf {\bibinfo {volume}
  {817}},\ \bibinfo {pages} {136330} (\bibinfo {year} {2021})}\BibitemShut
  {NoStop}%
\bibitem [{\citenamefont {Tangphati}\ \emph
  {et~al.}(2021{\natexlab{a}})\citenamefont {Tangphati}, \citenamefont
  {Pradhan}, \citenamefont {Errehymy},\ and\ \citenamefont
  {Banerjee}}]{Tangphati:2021tcy}%
  \BibitemOpen
  \bibfield  {author} {\bibinfo {author} {\bibfnamefont {T.}~\bibnamefont
  {Tangphati}}, \bibinfo {author} {\bibfnamefont {A.}~\bibnamefont {Pradhan}},
  \bibinfo {author} {\bibfnamefont {A.}~\bibnamefont {Errehymy}}, \ and\
  \bibinfo {author} {\bibfnamefont {A.}~\bibnamefont {Banerjee}},\ }\href
  {\doibase https://doi.org/10.1016/j.physletb.2021.136423} {\bibfield
  {journal} {\bibinfo  {journal} {Phys. Lett. B}\ }\textbf {\bibinfo {volume}
  {819}},\ \bibinfo {pages} {136423} (\bibinfo {year}
  {2021}{\natexlab{a}})}\BibitemShut {NoStop}%
\bibitem [{\citenamefont {Tangphati}\ \emph
  {et~al.}(2021{\natexlab{b}})\citenamefont {Tangphati}, \citenamefont
  {Pradhan}, \citenamefont {Errehymy},\ and\ \citenamefont
  {Banerjee}}]{Tangphati:2021mvu}%
  \BibitemOpen
  \bibfield  {author} {\bibinfo {author} {\bibfnamefont {T.}~\bibnamefont
  {Tangphati}}, \bibinfo {author} {\bibfnamefont {A.}~\bibnamefont {Pradhan}},
  \bibinfo {author} {\bibfnamefont {A.}~\bibnamefont {Errehymy}}, \ and\
  \bibinfo {author} {\bibfnamefont {A.}~\bibnamefont {Banerjee}},\ }\href
  {\doibase https://doi.org/10.1016/j.aop.2021.168498} {\bibfield  {journal}
  {\bibinfo  {journal} {Annals Phys.}\ }\textbf {\bibinfo {volume} {430}},\
  \bibinfo {pages} {168498} (\bibinfo {year} {2021}{\natexlab{b}})}\BibitemShut
  {NoStop}%
\bibitem [{\citenamefont {Burgio}\ and\ \citenamefont
  {Fantina}(2018)}]{Burgio:2018mcr}%
  \BibitemOpen
  \bibfield  {author} {\bibinfo {author} {\bibfnamefont {G.~F.}\ \bibnamefont
  {Burgio}}\ and\ \bibinfo {author} {\bibfnamefont {A.~F.}\ \bibnamefont
  {Fantina}},\ }\href {\doibase https://doi.org/10.1007/978-3-319-97616-7_6}
  {\bibfield  {journal} {\bibinfo  {journal} {Astrophys. Space Sci. Libr.}\
  }\textbf {\bibinfo {volume} {457}},\ \bibinfo {pages} {255} (\bibinfo {year}
  {2018})}\BibitemShut {NoStop}%
\bibitem [{\citenamefont {Blaschke}\ and\ \citenamefont
  {Chamel}(2018)}]{Blaschke:2018mqw}%
  \BibitemOpen
  \bibfield  {author} {\bibinfo {author} {\bibfnamefont {D.}~\bibnamefont
  {Blaschke}}\ and\ \bibinfo {author} {\bibfnamefont {N.}~\bibnamefont
  {Chamel}},\ }\href {\doibase https://doi.org/10.1007/978-3-319-97616-7\_7}
  {\bibfield  {journal} {\bibinfo  {journal} {Astrophys. Space Sci. Libr.}\
  }\textbf {\bibinfo {volume} {457}},\ \bibinfo {pages} {337} (\bibinfo {year}
  {2018})}\BibitemShut {NoStop}%
\bibitem [{\citenamefont {Beringer}\ \emph {et~al.}(2012)\citenamefont
  {Beringer} \emph {et~al.}}]{Beringer:2012}%
  \BibitemOpen
  \bibfield  {author} {\bibinfo {author} {\bibfnamefont {J.}~\bibnamefont
  {Beringer}} \emph {et~al.},\ }\href {\doibase 10.1103/PhysRevD.86.010001}
  {\bibfield  {journal} {\bibinfo  {journal} {Phys. Rev. D}\ }\textbf {\bibinfo
  {volume} {86}},\ \bibinfo {pages} {010001} (\bibinfo {year}
  {2012})}\BibitemShut {NoStop}%
\bibitem [{\citenamefont {Astashenok}\ \emph {et~al.}(2015)\citenamefont
  {Astashenok}, \citenamefont {Capozziello},\ and\ \citenamefont
  {Odintsov}}]{Astashenok2015}%
  \BibitemOpen
  \bibfield  {author} {\bibinfo {author} {\bibfnamefont {A.~V.}\ \bibnamefont
  {Astashenok}}, \bibinfo {author} {\bibfnamefont {S.}~\bibnamefont
  {Capozziello}}, \ and\ \bibinfo {author} {\bibfnamefont {S.~D.}\ \bibnamefont
  {Odintsov}},\ }\href {\doibase
  https://doi.org/10.1016/j.physletb.2015.01.030} {\bibfield  {journal}
  {\bibinfo  {journal} {Phys. Lett. B}\ }\textbf {\bibinfo {volume} {742}},\
  \bibinfo {pages} {160} (\bibinfo {year} {2015})}\BibitemShut {NoStop}%
\bibitem [{\citenamefont {Astashenok}\ \emph {et~al.}(2017)\citenamefont
  {Astashenok}, \citenamefont {Odintsov},\ and\ \citenamefont {de~la
  Cruz-Dombriz}}]{Astashenok2017}%
  \BibitemOpen
  \bibfield  {author} {\bibinfo {author} {\bibfnamefont {A.~V.}\ \bibnamefont
  {Astashenok}}, \bibinfo {author} {\bibfnamefont {S.~D.}\ \bibnamefont
  {Odintsov}}, \ and\ \bibinfo {author} {\bibfnamefont {A.}~\bibnamefont {de~la
  Cruz-Dombriz}},\ }\href {\doibase https://doi.org/10.1088/1361-6382/aa8971}
  {\bibfield  {journal} {\bibinfo  {journal} {Class. Quantum Grav.}\ }\textbf
  {\bibinfo {volume} {34}},\ \bibinfo {pages} {205008} (\bibinfo {year}
  {2017})}\BibitemShut {NoStop}%
\bibitem [{\citenamefont {Pretel}\ and\ \citenamefont
  {Duarte}(2022)}]{Pretel2022CQG}%
  \BibitemOpen
  \bibfield  {author} {\bibinfo {author} {\bibfnamefont {J.~M.~Z.}\
  \bibnamefont {Pretel}}\ and\ \bibinfo {author} {\bibfnamefont {S.~B.}\
  \bibnamefont {Duarte}},\ }\href {\doibase
  https://doi.org/10.1088/1361-6382/ac7a88} {\bibfield  {journal} {\bibinfo
  {journal} {Class. Quantum Grav.}\ }\textbf {\bibinfo {volume} {39}},\
  \bibinfo {pages} {155003} (\bibinfo {year} {2022})}\BibitemShut {NoStop}%
\bibitem [{\citenamefont {Fisher}\ and\ \citenamefont
  {Carlson}(2019)}]{Fisher2019}%
  \BibitemOpen
  \bibfield  {author} {\bibinfo {author} {\bibfnamefont {S.~B.}\ \bibnamefont
  {Fisher}}\ and\ \bibinfo {author} {\bibfnamefont {E.~D.}\ \bibnamefont
  {Carlson}},\ }\href {\doibase 10.1103/PhysRevD.100.064059} {\bibfield
  {journal} {\bibinfo  {journal} {Phys. Rev. D}\ }\textbf {\bibinfo {volume}
  {100}},\ \bibinfo {pages} {064059} (\bibinfo {year} {2019})}\BibitemShut
  {NoStop}%
\bibitem [{\citenamefont {Harko}\ and\ \citenamefont
  {Moraes}(2020)}]{Harko2020PRD}%
  \BibitemOpen
  \bibfield  {author} {\bibinfo {author} {\bibfnamefont {T.}~\bibnamefont
  {Harko}}\ and\ \bibinfo {author} {\bibfnamefont {P.~H. R.~S.}\ \bibnamefont
  {Moraes}},\ }\href {\doibase 10.1103/PhysRevD.101.108501} {\bibfield
  {journal} {\bibinfo  {journal} {Phys. Rev. D}\ }\textbf {\bibinfo {volume}
  {101}},\ \bibinfo {pages} {108501} (\bibinfo {year} {2020})}\BibitemShut
  {NoStop}%
\end{thebibliography}
\end{document}